\documentclass[12pt]{article}
\usepackage[xdvi]{graphicx}
\usepackage{amssymb}
\newcommand{\bea}{\begin{eqnarray}}
\newcommand{\eea}{\end{eqnarray}}

\makeatletter
\@addtoreset{equation}{section}
\makeatother

\begin{document}
\begin{titlepage}
%%%%% PREPRINT NUMBERS %%%%%%
\begin{flushright}
OU-HET 646/2009
\end{flushright}

\vspace{25ex}

%%%%%%%%%%%%%%%%%%% TITLE %%%%%%%%%%%%%%%%%%
\begin{center}
{\Large\bf 
Mass ratio of W and Z bosons \\ \vspace{1ex}
in SU(5) gauge-Higgs unification
}
\end{center}
%%%%%%%%%%%%%%%% AUTHORS %%%%%%%%%%%%%%%%%%%%%%%

\vspace{1ex}

\begin{center}
{\large
Nobuhiro Uekusa
}
\end{center}
%%%%%%%%%%%%%%%%%%%%%%% AFFILIATION %%%%%%%%%%%%
\begin{center}
{\it Department of Physics, 
Osaka University \\
%1-1 Machikaneyama, 
Toyonaka, Osaka 560-0043
Japan} \\
\textit{E-mail}: uekusa@het.phys.sci.osaka-u.ac.jp
\end{center}

%%%%%%%%%%%%%%%%%% ABSTRACT %%%%%%%%%%%%%%%

\vspace{3ex}

\begin{abstract}

We study the mass ratio of $W$ and $Z$ bosons 
in an SU(5) gauge-Higgs unification model with the
left-right symmetry remaining locally for boundary conditions
and with no scalar fields except for the
extra-dimensional component of gauge bosons.
We find that there are
a variety of boundary conditions
to yield distinct values of the mass ratio.

\end{abstract}
\end{titlepage}

\section{Introduction}

The gauge-Higgs unification is a candidate
of electroweak symmetry breaking~ \cite{%
hm,Hatanaka:1998yp}.
The four-dimensional Higgs field is 
identified with a part of the extra-dimensional
component of gauge bosons in higher dimensions.
The bosonic ingredients of the four-dimensional system
are gauge bosons and
scalar fields so that potentially 
it may have a phase transition as in the Coleman-Weinberg model for the scalar quantum electrodynamics. 
Because the four-dimensional scalar fields are 
originally higher-dimensional gauge fields,
the mass terms are not locally generated but 
can appear as non-local effects via a Wilson line phase.
This Wilson line phase is dynamical as 
an Aharonov-Bohm phase, where the
possible phase transition is
dynamical.

If the gauge-Higgs unification is a scenario
beyond the standard model, 
a simple setup is a model in which 
the Higgs doublet for SU(2)${}_L$ is included
as a part 
of the extra-dimensional component of higher-dimensional
gauge fields.
In the standard model, the Higgs sector
has the custodial symmetry SU(2)${}_D$.
In order that this property is taken into account
in the gauge-Higgs unification,
the left-right symmetry
SU(2)${}_L\times$SU(2)${}_R$ might be required for
compatibility 
with experimental data.
Here, it would be simple to 
treat such a left-right symmetry as a gauge symmetry.
A gauge-Higgs unification model 
as an extension of the electroweak standard model can 
be a gauge theory with 
a group larger than SU(2)${}_L\times$SU(2)${}_R$.
The large group should be broken to
U(1) for photon by
a Wilson line phase and by boundary conditions
with respect to extra dimensions.

It is nontrivial whether
the mass ratio of $W$ and $Z$ bosons
is correctly produced in a specified model
even when
an algebraic structure such as groups and multiplets
is prepared.
The standard model 
with a scalar field in the fundamental representation 
has degrees of freedom to adjust the mass ratio.
For no scalar fields  except for the extra-dimensional
component of higher-dimensional gauge fields
in the gauge-Higgs unification,
it should be clarified what the value of  
the mass ratio of $W$ and $Z$ bosons can be.

In this paper, we 
study the mass ratio of $W$ and $Z$ bosons 
in an SU(5) gauge-Higgs unification model
in the Randall-Sundrum warped spacetime.
The group SU(5) is broken by boundary conditions 
consistently
with higher-dimensional gauge transformation
to SO(5) at one boundary
and to [SU(2)$\times$U(1)]${}_L\times$ 
[SU(2)$\times$U(1)]${}_R$ at the other boundary.
Locally left-right symmetry remains.
The group with the overlapping of Neumann condition 
for the two boundary conditions is SU(2)$\times$U(1).
When a nonzero Wilson line phase is developed, 
the SU(2)$\times$U(1) is broken 
to U(1).
The rank of the gauge group is reduced at the levels of
SU(5)$\to$SO(5)
and SU(2)$\times$U(1)$\to$U(1).
In this model, there are
no scalar fields except for the
extra-dimensional component of gauge bosons,
that are required for
the symmetry breaking of SU(5) to U(1).
We work in two cases of boundary conditions for the rank
reduction SU(5)$\to$SO(5).
One is on the TeV brane 
and the other is on the Planck brane.
We find that in the toy model there are deviations of the values 
of the mass ratio of $W$ and $Z$ masses from 
the experimental value.
On the other hand,
the values are obtained differently in the two cases.
This shows that
there are
a variety of boundary conditions
to yield distinct values of the mass ratio,
%even with the same starting group and with the same
%subgroups of symmetry breaking,
as a room to build a realistic model for symmetry breaking
without additional scalar
fields.
We will discuss 
possible prospects to improve the model.

The paper is organized as follows. 
In Sec.~\ref{sec:model}, 
we present
our model including fields and action integrals.
The mass ratios for the case with the
rank reductions on the Plank brane
and on the TeV brane are given in Sec.~\ref{sec:pl}
and Sec.~\ref{sec:tev}, respectively. 
We conclude in Sec.~\ref{sec:concl} with some remarks.
Details of calculations and
formulas are shown in appendices.

\section{Model \label{sec:model}}

The model is defined in the Randall-Sundrum warped spacetime
whose metric is given by~\cite{rs}
\bea
   ds^2 = e^{-2\sigma(y)}\eta_{\mu\nu} dx^\mu dx^\nu
     +dy^2 ,
\eea
where $\eta_{\mu\nu} =\textrm{diag}(-1,1,1,1)$,
$\sigma(y)=\sigma(y+2L)$, 
and $\sigma(y)=k|y|$ for $|y|\leq L$.
The fundamental region in the fifth dimension is given
by $0\leq y\leq L$. The Planck brane and the TeV brane are located
by $y=1$ and $y=L$, respectively.
The bulk region $0<y<L$ is an anti-de Sitter spacetime with
the cosmological constant $\Lambda=-6k^2$.

We consider an SU(5) gauge theory in the gauge-Higgs unification.
The SU(5) symmetry is broken to SO(5) on one boundary and
to [SU(2)$\times$U(1)]${}_L\times$[SU(2)$\times$U(1)]${}_R$ on the other boundary.
The symmetry at this stage is 
broken by boundary conditions consistently with 
the five-dimensional gauge transformation~%
\cite{Csaki:2003dt}-%Sakai:2006qi,Csaki:2008se,
\cite{Uekusa:2008ag}.
The SU(5) five-dimensional gauge fields are 
$A_M=\sum_{i=1}^{24} A_M^i T_i$. 
The generators are written as $T_i={\lambda_i\over 2}$ where
$i=1,\cdots,24$ and $\lambda_i$ is defined in a similar way to
Gell-Mann matrices for SU(3).\footnote{%
Matrix forms of generators are explicitly summarized in Ref.~%
\cite{Uekusa:2008ag}.
Some bases to represent SO(5) are given in
Appendix~\ref{app:omela}.}
The generator of 
[SU(2)$\times$U(1)]${}_L\times$[SU(2)$\times$U(1)]${}_R$ is given by
$T_1,T_2,T_3,T_8,T_{15},T_{22},T_{23},T_{24}$.
The generators of SO(5) are given by
\bea
  && T_{\overline{1}}=\textrm{${1\over \sqrt{2}}$}
  (T_1 -T_{22})  ,
\quad 
   T_{\overline{2}}=\textrm{${1\over \sqrt{2}}$}
   (T_2 +T_{23}), 
\quad
    T_{\overline{3}}=\textrm{${1\over \sqrt{2}}$}
  (T_4 -T_{13}) ,
\quad
   T_{\overline{4}}=\textrm{${1\over \sqrt{2}}$}
  (T_5 -T_{14}) ,   
\nonumber
\\
 &&   T_{\overline{5}}=\textrm{${1\over \sqrt{2}}$}
  (T_6 -T_{20}),
 \quad
       T_{\overline{6}}=\textrm{${1\over \sqrt{2}}$}
  (T_7 -T_{21}) ,
  \quad
    T_{\overline{7}}=\textrm{${1\over \sqrt{2}}$}
  (T_{11} -T_{16})  ,
  \quad
        T_{\overline{8}}=\textrm{${1\over \sqrt{2}}$}
  (T_{12} -T_{17}) ,
\nonumber
\\
  &&   T_{\overline{9}}
    = \textrm{${1\over \sqrt{2}}$}(T_3 +\textrm{${\sqrt{6}\over 4}$}T_{15}
      -\textrm{${\sqrt{10}\over 4}$}T_{24})  ,
  \quad
  T_{\overline{10}}
    = \textrm{${1\over \sqrt{2}}$}(\textrm{${\sqrt{3}\over 3}$}
   T_8 +\textrm{${5\sqrt{6}\over 12}$}T_{15}
      +\textrm{${\sqrt{10}\over 4}$}T_{24})  .
  \label{so5l}
\eea
The overlapping generators of 
[SU(2)$\times$U(1)]${}_L\times$[SU(2)$\times$U(1)]${}_R$
and SO(5) are
$T_{\overline{1}}$,
$T_{\overline{2}}$,
$T_{\overline{9}}$,
$T_{\overline{10}}$ which form the
SU(2)$\times$U(1) algebra. 
The SU(2)$\times$U(1) part of the gauge fields
are written as
\bea
  A_\mu^{\overline{1}}T_{\overline{1}}
   +A_\mu^{\overline{2}}T_{\overline{2}}
   +A_\mu^{\overline{9}}T_{\overline{9}}
   +A_\mu^{\overline{10}}T_{\overline{10}}
   ={1\over 2}\left(\begin{array}{ccc}
    U_\mu && \\
      &0& \\
     && -U_\mu \\
     \end{array}\right). 
     \label{uvec}
\eea
Here 
$U_\mu ={1\over \sqrt{2}}
(W_\mu^a \sigma^a +B_\mu {\mathbf 1}_2)$,
with
$W_\mu^1 \equiv A_\mu^{\overline{1}}$,
$W_{\mu}^2 \equiv A_\mu^{\overline{2}}$,
$W_{\mu}^3 \equiv A_\mu^{\overline{9}}$,
$B_\mu \equiv A_\mu^{\overline{10}}$.
Hereafter we will call the boundary for SU(5)$\to$%
[SU(2)$\times$U(1)]${}_L\times$[SU(2)$\times$U(1)]${}_R$ 
the left-right symmetric boundary
and the boundary for SU(5)$\to$SO(5) 
the rank-reducing boundary.
At the left-right symmetric boundary,
the four-dimensional scalar fields, $A_y$ have Neumann boundary
condition for
$T_{4},\cdots, T_7$,
$T_9,\cdots , T_{14}$,
$T_{16}, \cdots, T_{21}$.
At the rank-reducing boundary,
$A_y$ have Neumann boundary condition
for
$T_{\overline{11}},\cdots, T_{\overline{20}}$,
$T_9,T_{10}, T_{18},T_{19}$.
The overlapping Neumann condition of both boundaries are given
by $T_{\overline{13}},\cdots, T_{\overline{18}}$,
$T_9,T_{10},T_{18},T_{19}$.
The Higgs doublet $H$ is included as 
\bea
    \sum_{\overline{i}=\overline{13}}^{\overline{16}}
   A_y^{\overline{i}} T_{\overline{i}}
  = {1\over 2}\left(\begin{array}{ccc}
     &H& \\
     H^\dag && H^T \\
      & H^* & \\
      \end{array}\right) ,\qquad
       H={1\over \sqrt{2}} 
   \left(\begin{array}{c}
    A_y^{\overline{13}}-iA_y^{\overline{14}} \\
    A_y^{\overline{15}} -iA_y^{\overline{16}} \\
    \end{array}\right) .
\eea

In identifying the spectrum of particles and their wave functions,
the conformal coordinate $z=e^{\sigma(y)}$ for the fifth dimension
is useful, with which the metric becomes
\bea
  ds^2 = {1\over z^2}
    \left\{ \eta_{\mu\nu} dx^\mu dx^\nu + 
      {dz^2\over k^2} \right\} ,\qquad
      z=e^{ky} .
\eea
The fundamental region $0\leq y\leq L$ is mapped
to $1\leq z \leq z_L =e^{kL}$.
Here $\partial_y =kz\partial_z$ and $A_y =kz A_z$.
The SU(5) gauge fields are split into classical and 
quantum parts $A_M=A_M^c +A_M^q$,
where $A_\mu^c=0$ and $A_y^c =kz A_z^c$.
The quadratic action for the SU(5) gauge fields is
\bea
   S_{\textrm{\scriptsize bulk 2}}^{%
  \textrm{\scriptsize gauge}} =
   \int d^4 x {dz\over kz}
  \left[ \textrm{tr}\left\{
   \eta^{\mu\nu} A_\mu^q (\Box + k^2 {\cal P}_4) A_\nu^q
   +k^2 A_z^q (\Box +k^2 {\cal P}_z)A_z^q 
   \right\} \right] ,
    \label{action}
\eea
for a gauge $\xi=1$ analogous to the
't Hooft-Feynman gauge in four dimensions. 
The differential operators are expressed as
$\Box = \eta^{\mu\nu}\partial_\mu \partial_\nu$,
${\cal P}_4= z{\cal D}_z^c ({1/ z}) {\cal D}_z^c$,
${\cal P}_z = {\cal D}_z^c z {\cal D}_z^c ({1/ z})$,
where ${\cal D}_M^c A_N^q =
\partial_M A_N^q - ig_A [ A_M^c, A_N^q]$.  
The $z$-dependence of the Higgs mode are given by
$\sqrt{2/(k(z_L^2 -1))} z$.

The Wilson line phase $\theta_H$ is given by
$\exp \left\{
i\theta_H (2 T_{\overline{15}})
    \right\}
  =\exp \left\{
ig_A \int_1^{z_L} dz \,\langle A_z \rangle 
\right\}$
so that
$\theta_H ={1\over 2} g_A v
\sqrt{(z_L^2 -1)/(2k)}$.
With a gauge transformation, a new basis in which 
the background field vanishes,
$\tilde{A}_z^c=0$ can be taken~\cite{%
Falkowski:2006vi,Hosotani:2007qw}.
In the twisted basis, gauge fields are defined as        
$\tilde{A}_M =\Omega A_M^q \Omega^{-1}$.
Here  $\Omega(z) =\exp \left\{
ig_A \int_z^{z_L} dz \,\langle A_z \rangle 
\right\}=\exp
\left\{ i\theta(z) (2T_{\overline{15}})
\right\}$ and
$\theta(z) =\theta_H 
(z_L^2 -z^2)/(z_L^2 -1)$.
The twist matrix $\Omega$ is written in a matrix form as
\bea
    \Omega(z) =
    \left(\begin{array}{ccccc}
     1& 0& 0& 0& 0 \\
     0& {c+1\over 2} & {is\over \sqrt{2}}
      & 0& {c-1 \over 2} \\
     0 & {is\over \sqrt{2}} & c &
     0& {is\over \sqrt{2}} \\
     0 & 0& 0& 1& 0 \\
     0 & {c-1\over 2} &
     {is\over \sqrt{2}} & 0& {c+1\over 2}\\
     \end{array}\right) ,
    \qquad \textrm{with}~~
    \left\{ \begin{array}{c}
   s\equiv \sin (\theta(z)) , \\
  c\equiv  \cos (\theta(z)) .\\
    \end{array}\right. 
\eea
For $\theta_H=0$, $\Omega(z)={\bf 1}_5$.
The twist matrix is unitary, $\Omega^{-1} =\Omega^\dag$.
The SU(5) gauge bosons are written in
the basis in terms of $T_i$ and in 
the basis with $T_{\overline{i}}$
as
$A_M^q = \sum_{i=1}^{24} A_M^{qi} T_i$
and   
$A_M^q=\sum_{\overline{i}=\overline{1}}^{\overline{20}}
     A_M^{q\overline{i}} T_{\overline{i}}
      +A_M^{q9}T_9 +A_M^{q10}T_{10} 
  +A_M^{q18}T_{18} +A_M^{q19} T_{19}$,
respectively.
The twisted fields are given in the same way by      
$\tilde{A}_M = \sum_{i=1}^{24} \tilde{A}_M^{qi} T_i
   =\sum_{\overline{i}=\overline{1}}^{\overline{20}}
     \tilde{A}_M^{q\overline{i}} T_{\overline{i}}
      +\tilde{A}_M^{q9}T_9 +\tilde{A}_M^{q10}T_{10} 
    +\tilde{A}_M^{q18}T_{18} +\tilde{A}_M^{q19} T_{19}$.
The component fields are related to each other under the basis transformations as
$A_M^{q1}\lambda_1 
   +A_M^{q2}\lambda_2 +\cdots
    =\Omega^{-1} (\tilde{A}_M^1 \lambda_1
     +\tilde{A}_M^2 \lambda_2 +\cdots)\Omega$
and
$A_M^{q\overline{1}} \lambda_{\overline{1}}
   +A_M^{q\overline{2}} \lambda_{\overline{2}}
   +\cdots
   =\Omega^{-1}
    (\tilde{A}_M^{\overline{1}} \lambda_{\overline{1}}
    +\tilde{A}_M^{\overline{2}} \lambda_{\overline{2}}
    +\cdots) \Omega$.
The mixing of generators with the twist 
matrix $\Omega$ is summarized in Appendix~\ref{%
app:omela}.
Basis transformations for gauge fields
can be read from the mixing of generators with $\Omega$.
The relations between component fields for 
$\tilde{A}_M^i \leftrightarrow A_M^{qi}$,
$\tilde{A}_M^{\overline{i}} \leftrightarrow A_M^{q\overline{i}}$,
$\tilde{A}_M^{\overline{i}}\leftrightarrow A_M^{i}$,
$\tilde{A}_M^i\leftrightarrow A_M^{q\overline{i}}$
are given in Appendix~\ref{ap:base}.

The twisted fields are decomposed
into four-dimensional and extra-dimensional parts
with mode functions as
\bea
  \tilde{A}_\mu^a (x,z) = \sum_n
   h_{A,n}^a (z) A_\mu^{(n)} (x) .
    \label{decom}
\eea
Here $n$ includes classification for $W$, $Z$ and photon.
From Eq.~(\ref{action}), the bulk equations for twisted fields 
force the mode functions
of the gauge bosons to
$h_{A,n}^a (z) \propto zJ_1(\lambda_n z)$, $zY_1 (\lambda_n z)$.
The mass eigenvalues are given by $k\lambda_n$ where
the subscript $n$ will be omitted
if no confusion arises. 
At $y=L$, the twist matrix is 
$\Omega(z_L)={\bf 1}_5$ with which $\tilde{A}_M =A_M^q$.
The basic function for Neumann condition at $y=0$ is
$C(z;\lambda) =
   {\pi \over 2}\lambda z z_L
     F_{1,0}(\lambda z, \lambda z_L)$.
The basic function for Dirichlet condition at $y=0$ is
$S(z;\lambda) =
 -{\pi\over 2} \lambda z F_{1,1} (\lambda z,
  \lambda z_L)$.
The functions $C(z;\lambda)$ and $S(z;\lambda)$ satisfy
the property $(S'C -C'S)(z;\lambda) = \lambda z$.
The other constants of the mode functions
will be determined by the boundary conditions
at $z=1$ and by the normalization.

The mass eigenvalue equations for gauge bosons
are obtained from the boundary conditions at $z=1$.
For the boundary conditions 
to yield symmetry breaking SU(5)$\to$SO(5)
and SU(5)$\to$[SU(2)$\times$U(1)]${}_L\times$
[SU(2)$\times$U(1)]${}_R$, the common
group SU(2)$\times$U(1) 
is assigned as Neumann condition for both boundaries.
The corresponding SU(2)$\times$U(1) gauge bosons would be massless
for vanishing Wilson line phase. 
When the Wilson line phase $\theta_H$ is nonzero,
a part of these gauge bosons acquire their masses.
The Wilson line phase is a dynamical phase and
it should be fixed dynamically.
On the other hand, the value $\theta_H$
depends on field contents.
Our interest is in possible values for 
the mass ratio of $W$ and $Z$ bosons
in the present model that has
less parameters to adjust.
As we will show explicitly,
the possibilities of allowed values of mass ratio 
are narrowed from the structure with respect to $\theta_H$
without knowing details of the dynamical fixing of $\theta_H$. 
We will not treat how the value of $\theta_H$ is
fixed further. 
We examine possible values of the ratio of the masses
by taking into account the $\theta_H$-dependence
of mass eigenvalue equations and the effective potential.

\section{Gauge bosons for 
Planck-brane rank reduction \label{sec:pl}} 

In this section, we analyze the mass ratio of gauge bosons
by placing the rank-reducing boundary for the Planck
brane and
the left-right symmetric boundary for the TeV brane.
The boundary conditions are given as
Neumann condition for 
$A_\mu^{q\overline{1}},\cdots,A_\mu^{q\overline{10}}$ 
at $y=0$
and Neumann condition for 
$A_\mu^{q1}$, $A_\mu^{q2}$,
$A_\mu^{q3}$, $A_\mu^{q8}$, $A_\mu^{q15}$,
$A_{\mu}^{q22}$, $A_{\mu}^{q23}$, $A_{\mu}^{q24}$ for $y=L$.

For the boundary conditions at $y=L$ where 
$\Omega(z_L)={\bf 1}_5$ and $\tilde{A}_M=A_M^q$, 
the mode functions have the forms
\bea
     h_{A,n}^i (z)=C_{A,n}^i C(z;\lambda_n) ,
  \qquad
     h_{A,n}^{\hat{i}} (z) =C_{A,n}^{\hat{i}} S(z;\lambda_n) ,
\eea     
where $i=1,2,3,8,15,22,23,24$ and
$\hat{i}=4,\cdots,7,9,\cdots,14,16,\cdots,21$.
The fundamental functions $C(z;\lambda_n)$ and $S(z;\lambda_n)$
are defined below Eq.~(\ref{decom})
and coefficients are denoted as $C_{A,n}^i$ 
and $C_{A,n}^{\hat{i}}$.
In the basis with $T_{\overline{i}}$,
the gauge bosons are written as
\bea
  \left(\begin{array}{c}
   A_\mu^{q\overline{1}} \\
   A_\mu^{q\overline{4}} \\
   A_\mu^{q\overline{7}} \\
   A_\mu^{q\overline{11}}\\
   A_\mu^{q\overline{14}} \\
   A_\mu^{q\overline{17}} \\
  \end{array}\right)
  &\!\!\!=\!\!\!&
  \left(\begin{array}{cccccc}
    {1 + c\over 2} & {s\over \sqrt{2}} &
    {1 - c\over 2} & {s\over \sqrt{2}} &
   {c-1 \over 2} & -{1 + c\over 2} \\
  -{s\over \sqrt{2}} &  c & 
  -{s \over \sqrt{2}} &  -c &
   -{s\over \sqrt{2}} & -{s\over \sqrt{2}} \\
 {1 - c \over 2} & -{s\over \sqrt{2}} &
  {1 + c \over 2} &  -{s\over \sqrt{2}} &
   -{1 + c\over 2} &  {c-1 \over 2} \\
   {1 + c\over 2} & {s\over \sqrt{2}} &
   {c-1 \over 2} & -{s\over \sqrt{2}} & 
   {c-1 \over 2} & {1 + c\over 2} \\
  -{s\over \sqrt{2}} & c & {s\over \sqrt{2}} & 
  c & -{s\over \sqrt{2}} &
   {s\over \sqrt{2}} \\
  {c-1\over 2} & {s\over \sqrt{2}} &
   {1 + c \over 2} & -{s\over \sqrt{2}} &
  {1 + c \over 2} & {c-1 \over 2}\\
  \end{array}\right) 
   \left(\begin{array}{c}
     C_{A,n}^1 C(z;\lambda_n) \\
     C_{A,n}^5 S(z;\lambda_n)\\
     C_{A,n}^{11} S(z;\lambda_n) \\
     C_{A,n}^{14} S(z;\lambda_n)\\
     C_{A,n}^{16} S(z;\lambda_n)\\
     C_{A,n}^{22} C(z;\lambda_n)\\
     \end{array}\right)
    {A_\mu^{(n)}(x) \over \sqrt{2}} ,
\nonumber
\eea    
for the components $(\overline{1},\overline{4},\overline{7},
\overline{11},\overline{14},\overline{17})$.
Here the components $(\overline{1},\overline{4},\overline{7},
\overline{11},\overline{14},\overline{17})$ for $A_M^q$
are linked to the components 
$(1,5,11,14,16,22)$ for $\tilde{A}_M$.
The mode expansion for 
all the other components is summarized
in Appendix~\ref{ap:gm}.
The components $(1,5,11,14,16,22)$ for $\tilde{A}_M$ are
a real part of the components for charged gauge bosons.
The boundary conditions at $y=0$ are
\bea
 0= \left(\begin{array}{cccccc}
    {1 + c\over 2}C' & {s\over \sqrt{2}}S' &
    {1 - c\over 2}S' & {s\over \sqrt{2}}S' &
   {c-1 \over 2}S' & -{1 + c\over 2}C' \\
  -{s\over \sqrt{2}} C'&  c S'& 
  -{s \over \sqrt{2}} S'&  -c S'&
   -{s\over \sqrt{2}} S'& -{s\over \sqrt{2}}C' \\
 {1 - c \over 2} C'& -{s\over \sqrt{2}} S'&
  {1 + c \over 2} S'&  -{s\over \sqrt{2}} S'&
   -{1 + c\over 2} S'&  {c-1 \over 2} C'\\
   {1 + c\over 2} C& {s\over \sqrt{2}} S&
   {c-1 \over 2} S& -{s\over \sqrt{2}} S& 
   {c-1 \over 2} S& {1 + c\over 2} C\\
  -{s\over \sqrt{2}} C& c S& {s\over \sqrt{2}} S& 
  c S& -{s\over \sqrt{2}} S&
   {s\over \sqrt{2}} C\\
  {c-1\over 2} C& {s\over \sqrt{2}} S&
   {1 + c \over 2} S& -{s\over \sqrt{2}} S&
  {1 + c \over 2} S& {c-1 \over 2}C\\
  \end{array}\right) 
   \left(\begin{array}{c}
     C_{A,n}^1 \\
     C_{A,n}^5 \\
     C_{A,n}^{11} \\
     C_{A,n}^{14} \\
     C_{A,n}^{16} \\
     C_{A,n}^{22} \\
     \end{array}\right)
  . \label{w1pl}
\eea
Here some abbreviations have been employed for
$C=C(z=1;\lambda_n)$,
$S=S(z=1;\lambda_n)$,
$C'=(dC/dz)(z=1;\lambda_n)$ and
$S'=(dS/dz)(z=1;\lambda_n)$.
At $z=1$, $s=\sin\theta_H$ and $c=\cos\theta_H$.
The condition that the determinant of the matrix
in Eq.~(\ref{w1pl}) vanishes 
means
\bea 
  SS' \left[
   2C'S + (1+c^2) (S'C-C'S) \right]
    \left[2C'S + s^2 (S'C-C'S) \right] = 0.
     \label{chargep}
\eea
With the relation $(S'C-C'S)|_{z=1} =\lambda$,
this equation is
\bea 
  SS' \left[
   2C'S +(1+c^2) \lambda \right]
     \left[2C'S + s^2 \lambda \right] = 0.
    \label{eqww}
\eea
If the Wilson line phase were zero,
a zero mode exists for
the boundary condition $C'S =0$.
This mode corresponds to one real component of $W$ boson.
For a nonzero Wilson line phase, $W$ boson acquires a mass
determined by $2C'S + s^2 \lambda =0$.
The product of the basic functions
is approximated as
$2C'S \approx -\lambda^3 kL e^{2kL}$ for $\lambda z_L \ll 1$ 
so that $W$ boson has the mass 
$m_W=k\lambda_W \approx\sqrt{k/L}e^{-kL}|s|$.
The mass depends on $\theta_H$ via a sine function.
From Eq.~(\ref{eqww}),
the components $(1,5,11,14,16,22)$ for $\tilde{A}_M$ 
include the mode with the boundary condition
$2CS' + (1+c^2) \lambda =0$.
This mode has the mass 
$m_{W'} = k\lambda_{W'} \approx \sqrt{k/L}e^{-kL}\sqrt{1+c^2}$.
If $|s|$ is order ${\cal O}(0.1\sim 1)$, 
$m_{W'}$ can be
of the same order as $m_W$.
Then $m_{W'}$ needs to be made larger.
If $|s|$ is so small that
$m_W \ll m_{W'}$, at low energy only $W$ boson survives.
For a very small $\theta_H$, 
$W$ boson 
without extra fields is simply obtained.
The mode for $SS'=0$ corresponds to massive Kaluza-Klein (KK) mode.

For the components $(\overline{2},\overline{3},
\overline{8},\overline{12},\overline{13},
\overline{18})$ for $A_M^q$, the boundary conditions at $y=0$ are
\bea
 0 =
  \left( \begin{array}{cccccc}
  {1 + c\over 2} C'& -{s\over \sqrt{2}} S'&
  {1 - c\over 2} S'& -{s\over \sqrt{2}} S'&
    {c-1 \over 2} S'& {1 + c\over 2} C'\\
 {s\over \sqrt{2}} C'& c S'& {s\over \sqrt{2}} S'&
  -c S'& {s\over \sqrt{2}} S'&
   -{s\over \sqrt{2}} C'\\
  {1 - c\over 2} C'&  {s\over \sqrt{2}} S'&
  {1 + c\over 2} S'& {s\over \sqrt{2}} S'& 
  -{1 + c\over 2} S'& {1 - c \over 2} C'\\
  {1 + c \over 2} C& -{s\over \sqrt{2}} S&
   {c-1 \over 2} S& {s\over \sqrt{2}} S&
    {c-1\over 2} S& -{1 + c\over 2} C\\
  {s\over \sqrt{2}} C& c S& -{s\over \sqrt{2}} S& 
  c S& {s\over \sqrt{2}} S& {s\over \sqrt{2}} C\\
 {c-1 \over 2} C& -{s\over \sqrt{2}}S&
  {1 + c\over 2} S& {s \over \sqrt{2}} S&
   {1 + c\over 2} S& {1 - c\over 2} C\\
  \end{array}\right)
  \left(\begin{array}{c}
    C_{A,n}^2  \\
    C_{A,n}^4 \\
    C_{A,n}^{12} \\
    C_{A,n}^{13} \\
    C_{A,n}^{17} \\
    C_{A,n}^{23} \\
    \end{array}\right)  .
\eea
Here the components $(\overline{2},\overline{3},
\overline{8},\overline{12},\overline{13},
\overline{18})$ for $A_M^q$ are linked to the components
(2, 4, 12, 13, 17, 23) for $\tilde{A}_M$.
These components 
%$(\overline{2},\overline{3},
%\overline{8},\overline{12},\overline{13},
%\overline{18})$
give the other real component of $W$ boson.
The condition that the determinant vanishes 
is the same as Eq.~(\ref{chargep}).
The charged $W$ boson is composed of the components
$(1,5,11,14,16,22)$
and $(2, 4, 12, 13, 17, 23)$ for $\tilde{A}_M$.
The charged $W'$ boson is also composed of the components
$(1,5,11,14,16,22)$
and $(2, 4, 12, 13, 17, 23)$ for $\tilde{A}_M$.
For the components $(\overline{5},19,\overline{15})$,
the boundary conditions at $y=0$ are
\bea
   0 =
     \left(\begin{array}{ccc}
    c S'& \sqrt{2} s S'& -c S'\\
   -s S& \sqrt{2} c S& s S\\
      S& 0 & S \\
      \end{array}\right)
 \left(\begin{array}{c}
    C_{A,n}^6 \\
    C_{A,n}^{19} \\
    C_{A,n}^{20} \\
    \end{array}\right) .
\eea
Here the components $(\overline{5},19,\overline{15})$ for
$A_M^q$ are linked to the components (6,19,20)
for $\tilde{A}_M$.
The condition that the determinant vanishes means
$S'S^2 =0$. This corresponds to massive KK mode.
For the components (9,10) for $A_M^q$ and $\tilde{A}_M$,
the boundary conditions at $y=0$ are $S=0$ which are massive 
KK mode.

For the components
$(\overline{6},\overline{9},\overline{10},
\overline{16},\overline{19},\overline{20},18)$,
the boundary conditions at $y=0$ are
\bea
  0 = E   
    (C_{A,n}^3 ,
    C_{A,n}^7 ,
    C_{A,n}^8 ,
    C_{A,n}^{15} ,
    C_{A,n}^{18} ,
    C_{A,n}^{21} ,
    C_{A,n}^{24})^T . 
   \label{plpl}
\eea
Here the components $(\overline{6},\overline{9},\overline{10},
\overline{16},\overline{19},\overline{20},18)$
for $A_M^q$
are linked to the components
(3, 7, 8, 15, 18, 21, 24) for $\tilde{A}_M$
and
the definition of $E$ is given in
Appendix~\ref{ap:bcpl}.
The components (3, 7, 8, 15, 18, 21, 24) for $\tilde{A}_M$ are
the mode of neutral gauge bosons.
The condition that the determinant of the matrix $E$
vanishes gives
a complicated equation with $\sin\theta_H$ and $\cos\theta_H$.
We consider typical cases for $\theta_H$:
a small $|s|$ and $s=1$. 
For a small $|s|$, it is simple to obtain $W$ boson without
extra light fields.
The other case $s=1$ can be a typical 
value for the dynamical phase as described
in the following, although extra fields need to be made heavier.
In the present model, scalar fields appear
from only the extra-dimensional component of gauge bosons.
For zero $\theta_H$, the potential $V(\theta_H=0)$ vanishes.
Because the $\theta_H$ is a phase,
physics is periodic for $\theta_H\to\theta_H +2\pi$ so that
$V(0) =V(2\pi)=0$.
A typical phase transition with
a stationary point for a nonzero $s$
could have the potential minimum at $\theta_H={\pi/2}$.
An example of such a symmetry breaking 
was given in a realistic model~\cite{%
Hosotani:2008tx}.
Then $s=1$.

For a small $|s|$ and $c=1-s^2/2+{\cal O}(s^4)$,
the condition $\det E=0$ is obtained as
\bea
0 =
  2gCC'S \left[C'S + s^2 {f\over 2g}\lambda\right]
  \left[ C'S +\lambda +s^2 \left({f-2b\over 2g}\right)
 \lambda\right] 
  +{\cal O}(s^4) ,
\eea
where
$g \equiv 6149 \sqrt{2} + 3177 \sqrt{5}$,
$f \equiv 773 \sqrt{2} + 539 \sqrt{5}$ and
$b \equiv 4603 \sqrt{2} + 2099 \sqrt{5}$.
The mode for $C'S=0$ is massless for the lowest level and
corresponds to photon.
The mode of $Z$ boson is obtained for 
$C'S + s^2 {f\over 2g}\lambda =0$.
The $Z$ boson mass is given by
$m_Z =k\lambda_{Z}
 \approx \sqrt{k/L}e^{-kL} |s| \sqrt{f/g}$.
Together with the mass of 
$W$ boson for the components
(1, 2, 4, 5, 11, $\cdots$, 14, 16, 17, 22, 23) for $\tilde{A}_M$,
this means the mass ratio
$m_W^2/m_Z^2 =g/f$.
Because $g/f >1$, for these boundary conditions and $|s|$,
$W$ boson is amount to be heavier than $Z$ boson.

For $s=1$, the condition $\textrm{det} E=0$ means
\bea
   0 = C C'S (C'S +\lambda)(13C'S +18\lambda) .
\eea
The mode for $C'S=0$ corresponds to photon.
The mode of $Z$ boson appears from
$C'S +\lambda=0$.
The mass is given by $m_Z =k\lambda_{\textrm{\scriptsize}}
\approx \sqrt{2k/L}e^{-kL}$.
The mass of the mode for $13C'S +18\lambda=0$
is $m_{Z'}=\sqrt{(36k)/(13L)} e^{-kL}$.
It needs to be made heavier.
Together with the mass of $W$ boson,
the mass ratio is given by
$m_W^2/m_Z^2 =1/2$ for $s=1$.
In this case,
$Z$ boson becomes heavier than $W$ boson
although there still remains a deviation of the mass ratio from
the experimental value.
As a characteristic aspect,
it has been found to depend on $\theta_H$ which boson is heavier 
of $Z$ boson and $W$ boson in the present context.
In next section,
it will be shown that the mass ratio
can change for the same value of $\theta_H$ 
and a simple exchange of the assignment of boundary conditions.

\section{Gauge bosons
for TeV-brane rank reduction \label{sec:tev}}

In this section, we analyze the masses of gauge bosons
interchanging the places of
the rank-reducing boundary and the left-right symmetric
boundary in Section~\ref{sec:pl}.
The boundary conditions are given as
Neumann condition for 
$A_{\mu}^{q1}$, $A_{\mu}^{q2}$,
$A_{\mu}^{q3}$, $A_{\mu}^{q8}$, $A_{\mu}^{q15}$,
$A_{\mu}^{q22}$, $A_{\mu}^{q23}$, $A_{\mu}^{q24}$ 
at $y=0$
and Neumann condition for 
$A_{\mu}^{q\overline{1}},\cdots,A_{\mu}^{q\overline{10}}$ 
for $y=L$.

For the boundary conditions at $y=L$ where 
$\Omega(z_L)={\bf 1}_5$ and $\tilde{A}_M=A_M^q$,
the mode functions have the forms
\bea
    h_{A,n}^i (z)= C_{A,n}^i C(z;\lambda_n) ,\qquad
    h_{A,n}^{\hat{i}} (z)=C_{A,n}^{\hat{i}}
      S(z;\lambda_n) ,
\eea
where $i=\overline{1},\cdots,\overline{10}$ and
$\hat{i}=\overline{11},\cdots,\overline{20},9,10,18,19$.
In the basis in terms of $T_i$, the gauge bosons are written
as
\bea
 \left(\begin{array}{c}
   A_\mu^{q1} \\
   A_\mu^{q5} \\
   A_\mu^{q11} \\
   A_\mu^{q14} \\
   A_\mu^{q16} \\
   A_\mu^{q22} \\
   \end{array}\right)
  &\!\!\! = \!\!\!&
   \left(\begin{array}{cccccc}
   {c +1\over 2} & {s\over \sqrt{2}} &
      -{c - 1\over 2}& {c + 1\over 2} 
    & {s\over \sqrt{2}}   &{c -1 \over 2} 
\\
 -{s\over \sqrt{2}} & c
   & {s\over \sqrt{2}} &  -{s\over \sqrt{2}}
  & c & -{s \over \sqrt{2}} 
\\
   -{c - 1\over 2} & {s\over \sqrt{2}} &
   {c + 1\over 2} &  {c-1 \over 2} 
  & -{s\over \sqrt{2}} & {c + 1\over 2}
\\
 -{s\over \sqrt{2}} & -c & {s\over \sqrt{2}} &
  {s\over \sqrt{2}} & c & {s\over \sqrt{2}}
\\
  {c-1 \over 2} & {s\over \sqrt{2}} &
   - {c+1 \over 2}  &  {c-1 \over 2} &
   {s\over \sqrt{2}} & {c + 1\over 2}
\\
 -{c + 1\over 2} & {s\over \sqrt{2}} &
   {c-1 \over 2} & {c + 1\over 2} &
  -{s\over \sqrt{2}} &  {c-1 \over 2}
\\   \end{array}\right)
    \left(\begin{array}{c}
     C_{A,n}^{\overline{1}} C(z;\lambda_n) \\
     C_{A,n}^{\overline{4}} C(z;\lambda_n)\\
     C_{A,n}^{\overline{7}} C(z;\lambda_n)\\
     C_{A,n}^{\overline{11}} S(z;\lambda_n)\\
     C_{A,n}^{\overline{14}} S(z;\lambda_n)\\
     C_{A,n}^{\overline{17}} S(z;\lambda_n)\\
  \end{array}\right)  {A_\mu^{(n)}(x) \over \sqrt{2}} ,
\nonumber
\eea
for the components (1,5,11,14,16,22).
The components (1,5,11,14,16,22) for $A_M^q$ 
are linked to the components 
$(\overline{1},\overline{4},\overline{7},
\overline{11},\overline{14},\overline{17})$ for
$\tilde{A}_M$.
The mode expansion for all the other components are
summarized in Appendix~\ref{ap:gm}.
The components $(\overline{1},\overline{4},\overline{7},
\overline{11},\overline{14},\overline{17})$ for
$\tilde{A}_M$ are
a real part of the components for charged gauge bosons.
The boundary conditions at $y=0$ are
\bea
   0=
   \left(\begin{array}{cccccc}
   {c +1\over 2} C' & {s\over \sqrt{2}} C' &
      -{c - 1\over 2} C' & {c + 1\over 2} S' 
    & {s\over \sqrt{2}} S'   &{c -1 \over 2} S'
\\
 -{s\over \sqrt{2}} C & c C
   & {s\over \sqrt{2}} C &  -{s\over \sqrt{2}} S
  & c S& -{s \over \sqrt{2}} S
\\
   -{c - 1\over 2} C & {s\over \sqrt{2}} C &
   {c + 1\over 2} C&  {c-1 \over 2} S 
  & -{s\over \sqrt{2}} S& {c + 1\over 2} S
\\
 -{s\over \sqrt{2}} C& -c C& {s\over \sqrt{2}}C &
  {s\over \sqrt{2}} S& c S& {s\over \sqrt{2}}S
\\
  {c-1 \over 2} C& {s\over \sqrt{2}} C&
   - {c+1 \over 2} C &  {c-1 \over 2} S&
   {s\over \sqrt{2}} S& {c + 1\over 2}S
\\
 -{c + 1\over 2} C'& {s\over \sqrt{2}} C'&
   {c-1 \over 2} C'& {c + 1\over 2} S'&
  -{s\over \sqrt{2}} S'&  {c-1 \over 2}S'
\\   \end{array}\right)
    \left(\begin{array}{c}
     C_{A,n}^{\overline{1}} \\
     C_{A,n}^{\overline{4}} \\
     C_{A,n}^{\overline{7}} \\
     C_{A,n}^{\overline{11}} \\
     C_{A,n}^{\overline{14}} \\
     C_{A,n}^{\overline{17}} \\
  \end{array}\right)   . 
   \label{wbc}
\eea
The condition that the determinant vanishes means
\bea
   CS \left[ (C'S + S'C)^2
  -c^4 (C'S - S'C)^2 \right] =0 .
\eea
With the relation $(S'C -C'S)|_{z=1} =\lambda$,
this equation is written as
\bea
   CS 
   \left[2C'S + (1+c^2) \lambda\right] 
 \left[2C'S + s^2 \lambda \right] =0.
   \label{eqwt}
\eea
As in Eq.~(\ref{eqww}),
the eigenvalue equation for $W$ boson is
$2C'S + s^2 \lambda =0$.
The mass of $W$ boson is 
$m_W=k\lambda_{W} \approx\sqrt{k/L}e^{-kL}|s|$.
Similarly to Eq.~(\ref{eqww}),
the components $(\overline{1},\overline{4},\overline{7},
\overline{11},\overline{14},\overline{17})$ for
$\tilde{A}_M$
include the mode with the boundary condition
$2C'S + (1+c^2) \lambda =0$.
For the components (2,4,12,13,17,23) for $A_M^q$,
the boundary conditions 
at $y=0$ are
\bea
  0 =
   \left(\begin{array}{cccccc}
   {c +1\over 2}C' & -{s\over \sqrt{2}} C'&
      -{c - 1\over 2}C'& {c + 1\over 2} S'
    & -{s\over \sqrt{2}} S'  &{c -1 \over 2}S' 
\\
 {s\over \sqrt{2}} C& c C
   & -{s\over \sqrt{2}} C&  {s\over \sqrt{2}} S
  & c S& {s \over \sqrt{2}} S
\\
   -{c - 1\over 2} C& -{s\over \sqrt{2}} C&
   {c + 1\over 2} C&  {c-1 \over 2} S
  & {s\over \sqrt{2}} S& {c + 1\over 2} S
\\
 {s\over \sqrt{2}} C& -c C& -{s\over \sqrt{2}} C&
  -{s\over \sqrt{2}} S& c S& -{s\over \sqrt{2}}S
\\
  {c-1 \over 2} C& -{s\over \sqrt{2}} C&
   - {c+1 \over 2} C &  {c-1 \over 2} S&
   -{s\over \sqrt{2}} S& {c + 1\over 2}S
\\
 {c + 1\over 2} C'& {s\over \sqrt{2}} C'&
   -{c-1 \over 2} C'& -{c + 1\over 2} S'&
  -{s\over \sqrt{2}} S'&  -{c-1 \over 2}S'
\\   \end{array}\right)
     \left(\begin{array}{c}
     C_{A,n}^{\overline{2}} \\
     C_{A,n}^{\overline{3}} \\
     C_{A,n}^{\overline{8}} \\
     C_{A,n}^{\overline{12}} \\
     C_{A,n}^{\overline{13}} \\
     C_{A,n}^{\overline{18}} \\
  \end{array}\right)   .
\eea
Here the components (2,4,12,13,17,23) for $A_M^q$
are linked to the components 
($\overline{2}$, $\overline{3}$, $\overline{8}$,
$\overline{12}$, $\overline{13}$, $\overline{18}$) for 
$\tilde{A}_M$.
The condition that the determinant vanishes is
the same as Eq.~(\ref{eqwt}).
The components ($\overline{2}$, $\overline{3}$, $\overline{8}$,
$\overline{12}$, $\overline{13}$, $\overline{18}$) for 
$\tilde{A}_M$
give the other real component
of $W$ boson.
For the components (6,19,20) for $A_M^q$, the boundary conditions 
at  $y=0$ are
\bea
  0  =
   \left(\begin{array}{ccc}
   c C& s S&  S\\
   -\sqrt{2} s  C& \sqrt{2} c S& 0 \\
   -c C& -s S& S \\
   \end{array}\right) 
   \left(\begin{array}{c}
    C_{A,n}^{\overline{5}}  \\
    C_{A,n}^{19}  \\
    C_{A,n}^{\overline{15}} \\
    \end{array}\right) .
\eea
Here the components (6,19,20) for $A_M^q$ are linked
to the components ($\overline{5}$, 19, $\overline{15}$)
for $\tilde{A}_M$.
The condition that the determinant vanishes means $CS^2=0$.
This corresponds to massive KK mode.
For the components (9,10) for $A_M^q$ and $\tilde{A}_M$,
the boundary conditions at $y=0$ are
$S=0$ which are massive KK modes.
For the components
(3,7,8,15,18,21,24) for $A_M^q$, the boundary conditions at $y=0$ are
\bea
 0 = L
     (C_{A,n}^{\overline{6}} ,
     C_{A,n}^{\overline{9}} ,
     C_{A,n}^{\overline{10}} ,
     C_{A,n}^{\overline{16}} ,
     C_{A,n}^{\overline{19}} ,
     C_{A,n}^{\overline{20}} ,
     C_{A,n}^{18})^T   .
  \label{pltev}
\eea
Here the components
(3,7,8,15,18,21,24) for $A_M^q$
are linked to the components 
($\overline{6}$, $\overline{9}$, $\overline{10}$,
$\overline{16}$, $\overline{19}$, $\overline{20}$) for
$\tilde{A}_M$ and
the definition of the matrix $L$ is given in
Appendix~\ref{ap:bcpl}.
The components ($\overline{6}$, $\overline{9}$, $\overline{10}$,
$\overline{16}$, $\overline{19}$, $\overline{20}$) for
$\tilde{A}_M$ are
the mode of neutral gauge bosons.
The condition that the determinant of $L$ vanishes gives
a complicated equation with $\sin\theta_H$ and $\cos\theta_H$.
Similarly to analysis in Section~\ref{sec:pl},
we consider typical cases for $\theta_H$:
a small $|s|$ and $s=1$. 
For a small $|s|$ and $c=1-{s^2/2}+{\cal O}(s^4)$,
the condition $\det L=0$ means
\bea
 0= C'S S' (16-5s^2)q
 \left[C'S 
  + \left(1+s^2 {21q-p\over 16q}\right)\lambda\right]
   \left[C'S +s^2 \lambda\right] 
  +{\cal O}(s^4) ,
\eea
where $q \equiv 383 \sqrt{2} + 156 \sqrt{5}$
and $p \equiv 24567 \sqrt{2} + 9884 \sqrt{5}$.
The mode for $C'S =0$ corresponds to photon.
The mode of $Z$ boson is for $C'S + s^2\lambda=0$.
The mass of $Z$ boson is 
$m_Z = k\lambda_{Z} 
\approx \sqrt{2k/L}e^{-kL}|s|$.
Together with the mass of $W$ boson, the mass ratio becomes
$m_W^2/m_Z^2=1/2$.
By comparing with the result in Section~\ref{sec:pl}, 
we find that the behavior of the mass ration 
significantly depends on the assignment of  
the place of the rank-reducing boundary.
For $s=1$, the condition 
$\textrm{det} L=0$ means $S' C'S [C'S +\lambda]^2=0$.
The mode of $C'S=0$ corresponds to photon.
The mass of $Z$ boson is 
$m_Z = k\lambda_{Z} 
\approx\sqrt{2k/L}e^{-kL}|s|$.
Then $m_W^2/m_Z^2 =1/2$.
For $s=1$,
the mass of $Z'$ boson is the same as the mass of $Z$ mass,
$m_{Z'}=m_Z$.
While in the present toy model 
there remains a deviation of the mass ration 
from the experimental value,
the mass ration has been found to change 
for the assignment of 
boundary conditions and the value of the Wilson line phase
that are essential elements in the 
gauge-Higgs unification.

Before closing the section, we would like to 
mention how gauge bosons are described with mode functions.
As an explicit example,
we focus on one real component of $W$ boson 
for the component $(\overline{1},\overline{4},\overline{7},
\overline{11},\overline{14},\overline{17})$ for
$\tilde{A}_M$. 
The boundary condition is given in Eq.~(\ref{wbc}). 
With the solution to eigenvectors for Eq.~(\ref{wbc}),
the gauge boson is written as
\bea
  {C_{A,n}^{\overline{1}} \over 1+c}
    W_\mu (x)
      \left[ (1+c) C(z;\lambda) T_{\overline{1}}
       +(1-c) C(z;\lambda) T_{\overline{7}}
       +\sqrt{2} s {C(z=1;\lambda)\over
         S(z=1;\lambda)} S (z;\lambda) T_{\overline{14}}
         \right] ,
         \label{wmodeh}
\eea
where $c=\cos \theta_H$ and $s=\sin \theta_H$.
The constant $C_{A,n}^{\overline{1}}$ is determined
by the normalization.
Details of a derivation are given in Appendix~\ref{%
ap:wmode}.
For zero $\theta_H$, $W$ boson is massless and has 
only the component for $T_{\overline{1}}$ which is a part 
of the generators of the corresponding unbroken SU(2)$\times$U(1)
as described in Eq.~(\ref{uvec}).
For nonzero $\theta_H$, $W$ boson acquires its mass
and is composed of the mixing of the three components 
$T_{\overline{1}}$, $T_{\overline{7}}$
and $T_{\overline{14}}$.
The other mode functions can be derived in a similar way.

\section{Conclusion \label{sec:concl}}

We have studied the mass ratio of $W$ and $Z$ bosons 
in an SU(5) gauge-Higgs unification model
with the Randall-Sundrum warped spacetime.
The group SU(5) is broken to SO(5) at one boundary and
[SU(2)$\times$U(1)]${}_L\times$
[SU(2)$\times$U(1)]${}_R$ 
at the other boundary
by boundary conditions.
The Higgs doublet for SU(2)${}_L$ is included as
the extra-dimensional component of five-dimensional
SU(5) gauge bosons.
For nonzero Wilson line phase, the overlapping 
SU(2)$\times$U(1) is broken to U(1).
Additional scalar fields are not required for the breaking
SU(5)$\to$U(1).
Because the starting group is a single group and
there are no additional scalar fields, 
possible values of the mass ratio for $W$ and $Z$ bosons 
are restrictive.
Instead of examining the potential for the Wilson line phase,
we have analyzed the mass ratio 
for typical values of the Wilson line phase $\theta_H$.

When the rank-reducing boundary where SU(5) is broken to SO(5)
is the Planck brane,
the mass ratio is $m_W^2/m_Z^2 =(6149\sqrt{2}+3177\sqrt{5})
/(773\sqrt{2}+539\sqrt{5})$ for $\theta_H \ll 1$ 
and $m_W^2/m_Z^2 =1/2$ for $\theta_H =\pi/2$.
When the rank-reducing boundary is the TeV brane,
the mass ratio is $m_W^2/m_Z^2=1/2$ for
both $\theta_H \ll 1$ and $\theta_H=\pi/2$.
For $\theta_H=\pi/2$, there are gauge bosons whose masses 
are of the same order as $m_W$ and $m_Z$.
These modes need to be made heavier.
While the values of the mass ratio in this toy model
deviate from the experimental value,
they have been found to depend on the 
assignment of the boundary conditions
even with the same starting group and with
the same pattern of subgroups for the symmetry breaking.
This gives a room to build a realistic model 
for symmetry breaking without additional scalar fields.

In order to produce the correct mass ratio,
we would like to discuss possible prospects to improve
the model.
While our starting point is a single group,
one might think that a direct product group is an 
alternative candidate.
However, that the starting point has direct product groups 
does not seem to be a favorable choice in the gauge-Higgs unification
because the mixing between groups as direct products
does not seem to occur via boundary conditions and via
the Wilson line phase without additional scalar fields.
We have found that for a fixed single group SU(5)
the assignment of boundary conditions can change the mass ratio.
A way to improve the mass ratio might
be to change a balance of the distribution for
the pattern of symmetry breaking.
As in the present model, a five-dimensional
spacetime has two endpoints for a finite 
extra-dimensional space.
Then the rank-reducing boundary and the left-right symmetric 
boundary are assigned piece by piece for each brane.
In six dimensions, there are four end-points with
respect to two finite extra-dimensional spaces.
If the left-right symmetric boundary 
is assigned for three branes 
and the rank-reducing boundary is assigned for the other single
brane, the resulting mass ratio could change
so as to respect left-right symmetry in a wider region
than in the five-dimensional case.
The two extra dimensions also involve
two dimensionful quantities.
The difference of the scales of the two extra dimensions 
can modify the values of low-energy physical quantities 
such as quark and charged lepton masses~\cite{%
Hall:2002qw,Uekusa:2008iz}.
Thus  the distribution for symmetry breaking 
and multiple dimensionful quantities might affect the 
mass ratio of $W$ and $Z$ bosons.
This analysis is left for future work.

\vspace{8ex}

%%%%%%%%%%%%%%%%%%%%
\subsubsection*{Acknowledgments}

This work is supported by Scientific Grants 
from the Ministry of Education
and Science, Grant No.~20244028.

\newpage

\begin{appendix}
 
\section{SU(5) generators and Wilson line mixing
\label{app:omela}}

\subsection{Base of SU(5) generators}

The $\lambda_i$ are represented like the
standard Gell-Mann matrices for
SU(3).
The matrices with barred indices such as
$\lambda_{\overline{i}}$ are defined as
\bea
   \left(\begin{array}{c}
    \lambda_{\overline{1}} \\
    \lambda_{\overline{11}} \\
    \end{array}\right)
  &\!\!\!=\!\!\!& G \left(\begin{array}{c}
      \lambda_1 \\
      \lambda_{22} \\
      \end{array}\right) ,
  \quad    
       \left(\begin{array}{c}
    \lambda_{\overline{2}} \\
    \lambda_{\overline{12}} \\
    \end{array}\right)
  = {1\over \sqrt{2}}
    \left(\begin{array}{cc}
      1 & 1 \\
      1 & -1 \\
      \end{array}\right) 
      \left(\begin{array}{c}
      \lambda_2 \\
      \lambda_{23} \\
      \end{array}\right) ,
\nonumber
\\      
%\eea
%\bea
   \left(\begin{array}{c}
    \lambda_{\overline{3}} \\
    \lambda_{\overline{13}} \\
    \end{array}\right)
  &\!\!\!=\!\!\!& G \left(\begin{array}{c}
      \lambda_4 \\
      \lambda_{13} \\
      \end{array}\right) ,
 \quad
       \left(\begin{array}{c}
    \lambda_{\overline{4}} \\
    \lambda_{\overline{14}} \\
    \end{array}\right)
  = G \left(\begin{array}{c}
      \lambda_5 \\
      \lambda_{14} \\
      \end{array}\right) ,
  \quad
   \left(\begin{array}{c}
    \lambda_{\overline{5}} \\
    \lambda_{\overline{15}} \\
    \end{array}\right)
  = G \left(\begin{array}{c}
      \lambda_6 \\
      \lambda_{20} \\
      \end{array}\right) ,
\nonumber
\\
%\eea
%\bea
       \left(\begin{array}{c}
    \lambda_{\overline{6}} \\
    \lambda_{\overline{16}} \\
    \end{array}\right)
  &\!\!\!=\!\!\!& G  \left(\begin{array}{c}
      \lambda_7 \\
      \lambda_{21} \\
      \end{array}\right) ,
      \quad
   \left(\begin{array}{c}
    \lambda_{\overline{7}} \\
    \lambda_{\overline{17}} \\
    \end{array}\right)
  = G \left(\begin{array}{c}
      \lambda_{11} \\
      \lambda_{16} \\
      \end{array}\right) ,
 \quad
       \left(\begin{array}{c}
    \lambda_{\overline{8}} \\
    \lambda_{\overline{18}} \\
    \end{array}\right)
  = G  \left(\begin{array}{c}
      \lambda_{12} \\
      \lambda_{17} \\
      \end{array}\right) ,
\nonumber
\\
%\eea
%\bea
  \left(\begin{array}{c}
    \lambda_{\overline{9}} \\
    \lambda_{\overline{10}} \\
    \lambda_{\overline{19}} \\
    \lambda_{\overline{20}} \\
    \end{array}\right)
  &\!\!\!=\!\!\!& {1\over \sqrt{2}}
  \left(\begin{array}{cccc}
  1 & 0 & {\sqrt{15}\over 4} &
  -{\sqrt{10}\over 4} \\
  0 & {\sqrt{3}\over 3} & {5\sqrt{6}\over 12}
  & {\sqrt{10}\over 4} \\
  1 & 0 & -{\sqrt{6}\over 4} & {\sqrt{10}\over 4} \\
  0 & {\sqrt{15}\over 3} &
  -{\sqrt{30}\over 12} & 
  -{\sqrt{2}\over 4} \\
  \end{array}\right) 
  \left(\begin{array}{c}
   \lambda_3 \\
   \lambda_8 \\
   \lambda_{15} \\
   \lambda_{24} \\
   \end{array}\right) ,
 \quad  \lambda_9, \lambda_{10} ,
   \lambda_{18}, \lambda_{19} .
\eea
where
$G ={1\over \sqrt{2}}
    \left(\begin{array}{cc}
      1 & -1 \\
      1 & 1 \\
      \end{array}\right)$.

When transformation matrices are denoted as ${\cal G}$ 
collectively,
the gauge bosons in a basis of matrices with barred indices are given by
\bea
  A_{\overline{j}} =\sum_i 
  ({\cal G}^T)_{\overline{j}i}  A_i ,\qquad
\textrm{with} \quad
  A_i \lambda_i =
   A_i {\cal G}_{i\overline{j}} 
   \lambda_{\overline{j}}
 = ({\cal G}^T)_{\overline{j}i}A_i \lambda_{\overline{j}} .
\eea
The relation between $A^{qi}$ and $A^{q\overline{i}}$ is
explicitly given as follows:
\bea
 \left(\begin{array}{c}
   A^{q1} \\
   A^{q5} \\
   A^{q11} \\
   A^{q14} \\
   A^{q16} \\
   A^{q22} \\
   \end{array}\right)
   &\!\!\!=\!\!\!&{1\over \sqrt{2}}
   \left(\begin{array}{cccccc}
   1 & 0 & 0 & 1 & 0 & 0 \\
   0 & 1 & 0 & 0 & 1 & 0 \\
   0 & 0 & 1 & 0 & 0 & 1 \\
   0 & -1 & 0 & 0 & 1 & 0 \\
   0 & 0& -1 & 0 & 0 & 1 \\
   -1 & 0 & 0 & 1 & 0 & 0 \\
   \end{array}\right)
   \left(\begin{array}{c}
    A^{q\overline{1}} \\
    A^{q\overline{4}} \\
    A^{q\overline{7}} \\
    A^{q\overline{11}} \\
    A^{q\overline{14}} \\
    A^{q\overline{17}} \\
    \end{array}\right) ,
\\
%\eea
%\bea
  \left(\begin{array}{c}
    A^{q2} \\
    A^{q4} \\
    A^{q12} \\
    A^{q13} \\
    A^{q17} \\
    A^{q23} \\
    \end{array}\right)
    &\!\!\!=\!\!\!&{1\over \sqrt{2}}
    \left(\begin{array}{cccccc}
    1 & 0 & 0 & 1 & 0 & 0 \\
    0 & 1 & 0 & 0 & 1 & 0 \\
    0 & 0 & 1 & 0 & 0 & 1 \\
    0 & -1 & 0 & 0 & 1 & 0 \\
    0 & 0& -1 & 0 & 0 & 1 \\
    1 & 0 & 0 &-1 & 0 &0 \\
    \end{array}\right)
    \left(\begin{array}{c}
    A^{q\overline{2}}\\
    A^{q\overline{3}} \\
    A^{q\overline{8}} \\
    A^{q\overline{12}}\\
    A^{q\overline{13}} \\
    A^{q\overline{18}} \\
    \end{array}\right) ,
\\    
%\eea
%\bea
  \left(\begin{array}{c}
   A^{q6} \\
   A^{q19} \\
   A^{q20} \\
   \end{array}\right)
  &\!\!\! =\!\!\!& {1\over \sqrt{2}}
   \left(\begin{array}{ccc}
   1 & 0 & 1 \\
   0 & \sqrt{2} & 0 \\
   -1 & 0 & 1 \\
   \end{array}\right) 
   \left(\begin{array}{c}
   A^{q\overline{5}} \\
   A^{q19} \\
   A^{q\overline{15}} \\
   \end{array}\right) ,
%\eea
%\bea
 \qquad  A^{q9} , A^{q10} ,
\\ 
%\eea
%\bea
  \left(\begin{array}{c}
    A^{q3} \\
    A^{q7} \\
    A^{q8} \\
    A^{q15} \\
    A^{q18} \\
    A^{q21} \\
    A^{q24} \\
    \end{array}\right) 
   &\!\!\! =\!\!\!& {1\over \sqrt{2}}
    \left(\begin{array}{ccccccc}
    0 & 1 & 0  & 0 & 1 & 0 & 0 \\
    1 & 0 & 0 & 1 & 0 & 0 & 0 \\
    0 & 0& {\sqrt{3}\over 3} & 0 &0& 
    {\sqrt{15}\over 3} & 0 \\
    0 & {\sqrt{15}\over 4} & {5\sqrt{6}\over 12} 
      & 0 & -{\sqrt{6}\over 4} &
      -{\sqrt{30}\over 12} & 0 \\
    0 & 0 &0 &0 &0 &0 & \sqrt{2} \\
    -1 & 0 & 0 & 1 & 0 & 0 &0 \\
    0 & -{\sqrt{10}\over 4} & {\sqrt{10}\over 4}
    & 0 & {\sqrt{10}\over 4} &
    -{\sqrt{2}\over 4} & 0 \\
    \end{array}\right) 
    \left(\begin{array}{c}
     A^{q\overline{6}} \\
     A^{q\overline{9}} \\
     A^{q\overline{10}} \\
     A^{q\overline{16}} \\
     A^{q\overline{19}} \\
     A^{q\overline{20}} \\
     A^{q18} \\
     \end{array}\right) .
\eea

\subsection{Mixing of SU(5) generators with $\Omega$}

\subsubsection{Mixing of $\lambda_i$ and $\Omega$}

The mixing of $\lambda_i$ with $\Omega$ is summarized as follows:
\bea
 \left(\begin{array}{c}
   \Omega^{-1} \lambda_1 \Omega \\
   \Omega^{-1} \lambda_5\Omega \\
    \Omega^{-1}\lambda_{16}\Omega \\
   \end{array}\right)
  &\!\!\!=\!\!\!&   
 M_1 \left(\begin{array}{c}
   \lambda_1  \\
   \lambda_5 \\ 
   \lambda_{16}  \\
   \end{array}\right) ,
 \qquad
 \left(\begin{array}{c}
 \Omega^{-1} \lambda_2 \Omega \\
 \Omega^{-1}\lambda_4 \Omega \\
 \Omega^{-1} \lambda_{17} \Omega \\
 \end{array}\right)
    =  M_1^T
  \left(\begin{array}{c}    
    \lambda_2 \\ 
    \lambda_4 \\
    \lambda_{17} \\
    \end{array}\right) ,
\\
%\eea
%\bea
   \left(\begin{array}{c}
  \Omega^{-1} \lambda_6 \Omega \\
  \Omega^{-1}\lambda_{19} \Omega \\
  \Omega^{-1} \lambda_{20} \Omega \\
   \end{array}\right)
  &\!\!\!=\!\!\!&
   M_2
  \left(\begin{array}{c}
     \lambda_6 \\ 
     \lambda_{19} \\
     \lambda_{20} \\
     \end{array}\right) ,
 \qquad
  \Omega^{-1} \lambda_9 \Omega 
  = \lambda_9 ,
  \qquad
  \Omega^{-1} \lambda_{10} \Omega = \lambda_{10} ,
\\
%\eea
%\bea
   \left(\begin{array}{c}
  \Omega^{-1} \lambda_{11}\Omega \\
  \Omega^{-1} \lambda_{14}\Omega \\
  \Omega^{-1}\lambda_{22}\Omega \\
  \end{array}\right)
    &\!\!\!=\!\!\!&    
  M_1^T
    \left(\begin{array}{c}
  \lambda_{11} \\
  \lambda_{14} \\
  \lambda_{22} \\
    \end{array}\right) ,
 \qquad
  \left(\begin{array}{c}
  \Omega^{-1} \lambda_{12} \Omega \\
  \Omega^{-1}\lambda_{13}\Omega \\
  \Omega^{-1} \lambda_{23} \Omega \\
  \end{array}\right)
   =  M_2
 \left(\begin{array}{c}
  \lambda_{12} \\
  \lambda_{13} \\
  \lambda_{23} \\
  \end{array}\right) ,
\\  
%\eea
%\bea     
 \left(\begin{array}{c}
  \Omega^{-1} \lambda_3 \Omega \\
  \Omega^{-1}\lambda_7 \Omega \\
  \Omega^{-1} \lambda_8 \Omega \\
  \Omega^{-1}\lambda_{15}\Omega \\
  \Omega^{-1} \lambda_{18}\Omega \\
  \Omega^{-1} \lambda_{21}\Omega \\
  \Omega^{-1} \lambda_{24}\Omega \\
  \end{array}\right)
  &\!\!\!=\!\!\!&
  V 
  \left(\begin{array}{c}
  \lambda_3  \\
  \lambda_7  \\
  \lambda_8  \\
  \lambda_{15} \\
  \lambda_{18} \\
  \lambda_{21} \\
  \lambda_{24} \\
  \end{array}\right) ,
  \qquad
  V= \left(\begin{array}{ccc}
     B_1  & F_1^T(-s) & B_2(s) \\
     F_1(s)  &  
   1+ {(c+3)(c-1)\over 48}
   & F_2(s) \\
   B_2^T(-s) & F_2^T(-s) & B_3 \\
   \end{array}\right) .
    \label{defv}
\eea
Here $c=\cos(\theta(z))$, $s=\sin(\theta(z))$ and
\bea 
 M_1 &\!\!\!=\!\!\!& \left(\begin{array}{ccc}
 {c+1\over 2} &
    -{s\over \sqrt{2}} &  {c-1\over 2} 
\\
 {s\over \sqrt{2}} &
    c & {s\over \sqrt{2}} 
\\
 {c-1\over 2} &
    -{s\over \sqrt{2}} &
    {c+1\over 2}  \\
    \end{array}\right) , 
     \qquad
  M_2 =
   \left(\begin{array}{ccc}
   {c+1\over 2} &
   -{s\over \sqrt{2}} &  
   -{c-1\over 2}  
\\  
   {s\over\sqrt{2}} &
      c  &
      -{s\over \sqrt{2}} 
\\
    -{c-1\over 2} &
     {s\over \sqrt{2}} &
     {c+1\over 2}
\\
 \end{array}\right) ,
\\
%\eea
%\bea
  B_1 &\!\!\!=\!\!\!& \left(\begin{array}{ccc}
  {1\over 2}\left\{
  c+1 +\left({c-1\over 2}\right)^2 \right\} 
  &
   {(c+1)s\over 2\sqrt{2}}
  &
     -{(5c+7)(c-1)\over 8\sqrt{3}}
\\
 -{(c+1)s\over 2\sqrt{2}} 
   &
    {(c+1)(2c-1)\over 2}
   &
    {(5c+1)s\over 2\sqrt{6}} 
\\
   -{(5c+7)(c-1)\over 8\sqrt{3}}
   &
    -{(5c+1)s\over 2\sqrt{6}}   
   &
   {25c^2 +2c-3\over 24} 
\\
  \end{array}\right) ,
\\
%\eea
%\bea
  B_2(s) &\!\!\!=\!\!\!& \left(\begin{array}{ccc}
    {s^2\over 4} 
  &  
     -{(c-1)s\over 2\sqrt{2}}
   &
    {\sqrt{10}(c-1)^2 \over 16} 
\\ 
   {cs\over \sqrt{2}} 
   &
     -{(c-1)(2c+1)\over 2}
   &  
    -{\sqrt{5}(c-1)s\over 4} 
\\
   -{5s^2 \over 4\sqrt{3}}
   &
  {(5c-1)s\over 2\sqrt{6}}
   &
   -{\sqrt{30}(5c+3)(c-1)\over 48} 
\\ 
    \end{array}\right) ,
\\    
%\eea
%\bea
   B_3 &\!\!\!=\!\!\!& \left(\begin{array}{ccc}
    {c^2 +1\over 2}
    &
     {cs\over \sqrt{2}}
     &
      {\sqrt{10}s^2\over 8}
\\
   -{cs\over \sqrt{2}} 
   &
    {(2c-1)(c+1)\over 2} 
    &
    {\sqrt{5}(c+1)s\over 4}
\\
   {\sqrt{10}s^2\over 8} 
   &
  -{\sqrt{5}(c+1)s\over 4} 
   &
    {5c^2+10c+1\over 16} \\
   \end{array}\right) ,
\\
%\eea
%\bea
 F_1(s) &\!\!\!=\!\!\!& \left({(c-1)^2 \over 8\sqrt{6}}  ,
  {(c-1)s\over 4\sqrt{3}}  ,
  -{(c-1)(5c+3)\over 24\sqrt{2}} \right) ,
\\
%\eea
%\bea
  F_2(s) &\!\!\!=\!\!\!& \left(
 {s^2\over 4\sqrt{6}} ,
  -{(c+1)s\over 4\sqrt{3}} ,
  {\sqrt{15}(c+3)(c-1)\over 48}\right) .
\eea
For $c=1$ and $s=0$, $V={\bf 1}_7$.

\subsubsection{Mixing of $\lambda_{\bar{i}}$ and 
$\Omega$}

The mixing of $\lambda_{\bar{i}}$ 
with $\Omega$ is summarized as follows:
\bea
   \left(\begin{array}{c}
   \Omega^{-1} \lambda_{\overline{1}}\Omega \\
   \Omega^{-1} \lambda_{\overline{4}}\Omega \\
   \Omega^{-1} \lambda_{\overline{7}}\Omega \\
   \Omega^{-1} \lambda_{\overline{11}}\Omega \\
   \Omega^{-1} \lambda_{\overline{14}}\Omega \\
   \Omega^{-1} \lambda_{\overline{17}}\Omega \\
   \end{array}\right)
   &\!\!\!=\!\!\!&
   \left(\begin{array}{cccccc}
   {c+1\over 2} &0 & -{c-1\over 2} & 0 &
   -{s\over \sqrt{2}} & 0\\
   0& c & 0 & {s\over \sqrt{2}} & 0 & {s\over \sqrt{2}} 
    \\
     -{c-1\over 2} &0 & {c+1\over 2} & 0 &
   {s\over \sqrt{2}} & 0\\
   0& -{s\over \sqrt{2}} & 0 &
   {c+1\over 2} & 0 & {c-1\over 2}  
    \\
     {s\over \sqrt{2}} &0 & -{s\over \sqrt{2}} & 0
     & c & 0\\
   0& -{s\over \sqrt{2}} & 0 
    & {c-1\over 2} & 0 & {c+1\over 2} 
    \\
  \end{array}\right)
   \left(\begin{array}{c}
     \lambda_{\overline{1}} \\
     \lambda_{\overline{4}} \\
     \lambda_{\overline{7}} \\
     \lambda_{\overline{11}} \\
     \lambda_{\overline{14}} \\
     \lambda_{\overline{17}} \\
  \end{array}\right)   ,
\\  
%\eea
%\bea
   \left(\begin{array}{c}
   \Omega^{-1} \lambda_{\overline{2}}\Omega \\
   \Omega^{-1} \lambda_{\overline{3}}\Omega \\
   \Omega^{-1} \lambda_{\overline{8}}\Omega \\
   \Omega^{-1} \lambda_{\overline{12}}\Omega \\
   \Omega^{-1} \lambda_{\overline{13}}\Omega \\
   \Omega^{-1} \lambda_{\overline{18}}\Omega \\
   \end{array}\right)
  &\!\!\! =\!\!\!&
   \left(\begin{array}{cccccc}
   {c+1\over 2} &0 & -{c-1\over 2} & 0 &
   {s\over \sqrt{2}} & 0\\
   0& c & 0 & -{s\over \sqrt{2}} & 0 & 
  -{s\over \sqrt{2}} 
    \\
     -{c-1\over 2} &0 & {c+1\over 2} & 0 &
   -{s\over \sqrt{2}} & 0\\
   0& {s\over \sqrt{2}} & 0 &
   {c+1\over 2} & 0 & {c-1\over 2}  
    \\
     -{s\over \sqrt{2}} &0 & {s\over \sqrt{2}} & 0
     & c & 0\\
   0& {s\over \sqrt{2}} & 0 
    & {c-1\over 2} & 0 & {c+1\over 2} 
    \\
  \end{array}\right)
   \left(\begin{array}{c}
     \lambda_{\overline{2}} \\
     \lambda_{\overline{3}} \\
     \lambda_{\overline{8}} \\
     \lambda_{\overline{12}} \\
     \lambda_{\overline{13}} \\
     \lambda_{\overline{18}} \\
  \end{array}\right)   ,
\\
%\eea
%\bea
  \left(\begin{array}{c}
    \Omega^{-1} \lambda_{\overline{5}}\Omega \\
    \Omega^{-1} \lambda_{19}\Omega \\
   \end{array}\right)
  &\!\!\!=\!\!\!& \left(\begin{array}{cc}
     c & -s \\
     s & c \\
     \end{array}\right)
     \left(\begin{array}{c}
     \lambda_{\overline{5}} \\
     \lambda_{19} \\
     \end{array}\right) ,
\qquad
 \left\{
  \begin{array}{c} 
 \Omega^{-1} \lambda_{\overline{15}}\Omega
 = \lambda_{\overline{15}} , \\
 \Omega^{-1}\lambda_9 \Omega =\lambda_9, \\
 \Omega^{-1} \lambda_{10}\Omega =\lambda_{10} ,
  \end{array} \right.
\\
%\eea
%\bea
   \left(\begin{array}{c}
   \Omega^{-1} \lambda_{\overline{6}}\Omega \\
   \Omega^{-1} \lambda_{\overline{9}}\Omega \\
   \Omega^{-1} \lambda_{\overline{10}}\Omega \\
   \Omega^{-1} \lambda_{\overline{16}}\Omega \\
   \Omega^{-1} \lambda_{\overline{19}}\Omega \\
   \Omega^{-1} \lambda_{\overline{20}}\Omega \\
   \Omega^{-1} \lambda_{18}\Omega \\
   \end{array}\right)
  &\!\!\! =\!\!\!& U
    \left(\begin{array}{c}
     \lambda_{\overline{6}} \\
     \lambda_{\overline{9}} \\
     \lambda_{\overline{10}} \\
     \lambda_{\overline{16}} \\
     \lambda_{\overline{19}} \\
     \lambda_{\overline{20}} \\
     \lambda_{18} \\
  \end{array}\right)   ,
  \qquad
    U = (U_1 ~ U_2) .
     \label{defu}
\eea
Here $c=\cos(\theta(z))$, $s=\sin(\theta(z))$ and
\bea
   U_1 &\!\!\!=\!\!\!&
   \left(\begin{array}{cccc}
       2c^2 - 1 & -{\sqrt{2}s(2+3c)\over 26+3\sqrt{10}} &
       {5\sqrt{2}s(-8+(14+3\sqrt{10})c)
     \over 12(26+3\sqrt{10})} & 0 
\\
       {(-\sqrt{2}+\sqrt{5})cs\over 16} &
  {8+4\sqrt{10}+(18-\sqrt{10})c
     \over 26+3\sqrt{10}} & -{(98+169\sqrt{10})(c-1)\over 1172}
   & {(9\sqrt{2}-\sqrt{5})s \over 16}
\\
       0 & -{16 (c-1)\over 26+3\sqrt{10}} &
 {18-\sqrt{10}+4(2+\sqrt{10})c\over
     26+3\sqrt{10}} 
  & -{s\over \sqrt{2}}
\\
       0 & {2\sqrt{2}s(c-9)\over 
     26+ 3\sqrt{10}} & {2\sqrt{2}s(7+6\sqrt{10}+5c)\over
     3(26+3\sqrt{10})} & c \\
       {cs\over \sqrt{2}} & 0 & 0 & 0\\
       -{\sqrt{10}cs\over 2} &0 & 0 &0\\
       -cs & 0 & 0 &0\\
       \end{array}\right) ,
\\
%\eea
%\bea
   U_2 &\!\!\!=\!\!\!&
   \left(\begin{array}{ccc}
   {s(4-(20+3\sqrt{10})c)\over 
   \sqrt{2}(26+3\sqrt{10})} &{s(4\sqrt{10}+75c +71\sqrt{10} c)
   \over 6(26+3\sqrt{10})} & cs 
 \\
   {(1-c)(149-70\sqrt{10}
   +(11-10\sqrt{10})c)
   \over 32(26 +3\sqrt{10})} & {(c-1)(50\sqrt{2} -35\sqrt{5}
   -50\sqrt{2} c +11\sqrt{5}c)
   \over 
   32(26+3\sqrt{10})} & {(-\sqrt{2}+\sqrt{5})s^2
   \over 32} 
\\
   -{3(-41+16\sqrt{10})(c-1)
   \over 586} &-{(80\sqrt{2} -41\sqrt{5})(c-1)
   \over 586} & 0
\\
   {s(10-3\sqrt{10}-4c)
   \over \sqrt{2}(26+3\sqrt{10})} &-{s(15-5\sqrt{10} +2\sqrt{10}c)
   \over 3(26+3\sqrt{10})} & 0
\\
   {c^2 +3\over 4} &-{\sqrt{5}(c^2-1)\over 4}
   & {\sqrt{2}s^2\over 4}  \\
   -{\sqrt{5}(c^2-1)\over 4} &{5c^2-1\over 4} 
   & -{\sqrt{10}s^2\over 4} \\
   {\sqrt{2}s^2\over 4} &-{\sqrt{10}s^2\over 4} & {c^2+1\over 2} \\
   \end{array}\right) .
\eea
For $c=1$ and $s=0$, $U={\bf 1}_7$.

\section{Base transformation between fields
\label{ap:base}}

In this appendix, the base transformations between 
fluctuation fields and twisted fields are given.
Fluctuation fields and twisted fields are related to 
each other through
$A_M^q= \Omega^{-1} \tilde{A}_M \Omega$.
All the relations for 
component fields are derived from this equation.
For $A_M^{qi}$ and $\tilde{A}_M^i$ which correspond to
generators $T_i$,
the component fields are included as $A_M^{qi}\lambda_i 
 = \Omega^{-1} \tilde{A}_M^i \lambda_i \Omega 
 =\tilde{A}_M^{i} \sum_j c_{ij}\lambda_j$
where $c_{ij}$ are the coefficients for the mixing of
$\lambda_i$ with $\Omega$.
With the same $c_{ij}$ being expressed as elements of a matrix,
the relations for components are given by
$A_M^{qj} =\sum_i \tilde{A}_M^i c_{ij} 
  =\sum_i (c^T)_{ji} \tilde{A}_M^i$. 
Explicit equations are summarized in the following.

\subsection{$A_M^{qi}\leftrightarrow \tilde{A}_M^i$}

\bea
   \left(\begin{array}{c}
   A_M^{q1} \\
   A_M^{q5} \\
   A_M^{q11} \\
   A_M^{q14} \\
   A_M^{q16} \\
   A_M^{q22} \\
   \end{array}\right)
   &\!\!\!=\!\!\!& \left(\begin{array}{cccccc}
   {c+1\over 2} & {s\over \sqrt{2}} &
   0& 0& {c-1\over 2}& 0\\
   -{s\over \sqrt{2}} & c& 0& 0& -{s\over \sqrt{2}}&
   0 \\
   0& 0& {c+1\over 2} & -{s\over \sqrt{2}}&
   0 & {c-1\over 2} \\
   0& 0& {s\over \sqrt{2}} & c& 0& {s\over \sqrt{2}} \\
   {c-1\over 2} & {s\over \sqrt{2}}&
   0& 0& {c+1\over 2}& 0 \\
   0& 0& {c-1\over 2} & -{s\over \sqrt{2}} &
   0& {c+1\over 2}  \\
   \end{array}\right) 
   \left(\begin{array}{c}
    \tilde{A}_M^1 \\
    \tilde{A}_M^5 \\
    \tilde{A}_M^{11} \\
    \tilde{A}_M^{14} \\
    \tilde{A}_M^{16} \\
    \tilde{A}_M^{22} \\
    \end{array}\right) ,
\\
%\eea
%\bea
   \left(\begin{array}{c}
    A_M^{q2} \\
    A_M^{q4} \\
    A_M^{q12} \\
    A_M^{q13} \\
    A_M^{q17} \\
    A_M^{q23} \\
    \end{array}\right)
   &\!\!\! =\!\!\!& \left(\begin{array}{cccccc}
    {c+1\over 2}& -{s\over \sqrt{2}} & 0& 0&
    {c-1\over 2}& 0 \\
    {s\over \sqrt{2}} & c& 0& 0& {s\over \sqrt{2}}&
    0 \\
    0& 0& {c+1\over 2}& {s\over \sqrt{2}} & 0&
    -{c-1\over 2} \\
    0& 0& -{s\over \sqrt{2}}& c& 0& {s\over \sqrt{2}} \\
    {c-1\over 2} & -{s\over \sqrt{2}}& 0& 0&
    {c+1\over 2} & 0 \\
    0& 0& -{c-1\over 2} & -{s\over \sqrt{2}}& 0& 
    {c+1\over 2} \\
    \end{array}\right)
    \left(\begin{array}{c}
    \tilde{A}_M^2 \\
    \tilde{A}_M^4 \\
    \tilde{A}_M^{12} \\
    \tilde{A}_M^{13} \\
    \tilde{A}_M^{17} \\
    \tilde{A}_M^{23} \\
    \end{array}\right) ,
\\
%\eea
%\bea
   \left(\begin{array}{c}
    A_M^{q6} \\
    A_M^{q19} \\
    A_M^{q20} \\
    \end{array}\right)
  &\!\!\!= \!\!\!& \left(\begin{array}{ccc}
    {c+1\over 2} & {s\over \sqrt{2}} &
    -{c-1\over 2} \\
    -{s\over \sqrt{2}} & c & {s\over \sqrt{2}} \\
    -{c-1\over 2} & -{s\over \sqrt{2}}&
    {c+1\over 2} \\
    \end{array}\right)
    \left(\begin{array}{c}
    \tilde{A}_M^6 \\
    \tilde{A}_M^{19} \\
    \tilde{A}_M^{20} \\
    \end{array}\right) ,
\quad
  \left(\begin{array}{c}
   A_M^{q3} \\
   A_M^{q7} \\
   A_M^{q8} \\
   A_M^{q15} \\
   A_M^{q18} \\
   A_M^{q21} \\
   A_M^{q24} \\
   \end{array}\right)
   =V^T
   \left(\begin{array}{c}
    \tilde{A}_M^3 \\
    \tilde{A}_M^7 \\
    \tilde{A}_M^8 \\
    \tilde{A}_M^{15} \\
    \tilde{A}_M^{18} \\
    \tilde{A}_M^{21} \\
    \tilde{A}_M^{24} \\
    \end{array}\right) ,
\eea
and $A_M^{q9}=\tilde{A}_M^9$, $A_M^{q10}=\tilde{A}_M^{10}$.
Here $V$ is given in Eq.~(\ref{defv}).

\subsection{$A_M^{q\overline{i}}\leftrightarrow 
\tilde{A}_M^{\overline{i}}$}

\bea
   \left(\begin{array}{c}
   A_M^{q\overline{1}} \\
   A_M^{q\overline{4}} \\
   A_M^{q\overline{7}} \\
   A_M^{q\overline{11}} \\
   A_M^{q\overline{14}} \\
   A_M^{q\overline{17}} \\
   \end{array}\right)
  &\!\!\! =\!\!\!&
   \left(\begin{array}{cccccc}
   {c+1\over 2} &0 & -{c-1\over 2} & 0 &
   -{s\over \sqrt{2}} & 0\\
   0& c & 0 & {s\over \sqrt{2}} & 0 & {s\over \sqrt{2}} 
    \\
     -{c-1\over 2} &0 & {c+1\over 2} & 0 &
   {s\over \sqrt{2}} & 0\\
   0& -{s\over \sqrt{2}} & 0 &
   {c+1\over 2} & 0 & {c-1\over 2}  
    \\
     {s\over \sqrt{2}} &0 & -{s\over \sqrt{2}} & 0
     & c & 0\\
   0& -{s\over \sqrt{2}} & 0 
    & {c-1\over 2} & 0 & {c+1\over 2} 
    \\
  \end{array}\right)^T
   \left(\begin{array}{c}
     \tilde{A}_M^{\overline{1}} \\
     \tilde{A}_M^{\overline{4}} \\
     \tilde{A}_M^{\overline{7}} \\
     \tilde{A}_M^{\overline{11}} \\
     \tilde{A}_M^{\overline{14}} \\
     \tilde{A}_M^{\overline{17}} \\
  \end{array}\right)   ,
\\
%\eea
%\bea
   \left(\begin{array}{c}
   A_M^{q\overline{2}} \\
   A_M^{q\overline{3}} \\
   A_M^{q\overline{8}} \\
   A_M^{q\overline{12}} \\
   A_M^{q\overline{13}} \\
   A_M^{q\overline{18}} \\
   \end{array}\right)
  &\!\!\! =\!\!\!&
   \left(\begin{array}{cccccc}
   {c+1\over 2} &0 & -{c-1\over 2} & 0 &
   {s\over \sqrt{2}} & 0\\
   0& c & 0 & -{s\over \sqrt{2}} & 0 & 
  -{s\over \sqrt{2}} 
    \\
     -{c-1\over 2} &0 & {c+1\over 2} & 0 &
   -{s\over \sqrt{2}} & 0\\
   0& {s\over \sqrt{2}} & 0 &
   {c+1\over 2} & 0 & {c-1\over 2}  
    \\
     -{s\over \sqrt{2}} &0 & {s\over \sqrt{2}} & 0
     & c & 0\\
   0& {s\over \sqrt{2}} & 0 
    & {c-1\over 2} & 0 & {c+1\over 2} 
    \\
  \end{array}\right)^T
   \left(\begin{array}{c}
     \tilde{A}_M^{\overline{2}} \\
     \tilde{A}_M^{\overline{3}} \\
     \tilde{A}_M^{\overline{8}} \\
     \tilde{A}_M^{\overline{12}} \\
     \tilde{A}_M^{\overline{13}} \\
     \tilde{A}_M^{\overline{18}} \\
  \end{array}\right)   ,
\\
%\eea
%\bea
   \left(\begin{array}{c}
   A_M^{q\overline{5}} \\
   A_M^{q 19} \\
   A_M^{q\overline{15}} \\
   \end{array}\right)
   &\!\!\!= \!\!\!&
   \left(\begin{array}{ccc}
   c & s & 0 \\
   -s & c & 0 \\
   0 & 0 & 1 \\
   \end{array}\right) 
   \left(\begin{array}{c}
    \tilde{A}_M^{\overline{5}} \\
    \tilde{A}_M^{19} \\
    \tilde{A}_M^{\overline{15}} \\
    \end{array}\right) ,
\qquad
   \left(\begin{array}{c}
   A_M^{q\overline{6}} \\
   A_M^{q\overline{9}} \\
   A_M^{q\overline{10}} \\
   A_M^{q\overline{16}} \\
   A_M^{q\overline{19}} \\
   A_M^{q\overline{20}} \\
   A_M^{q18} \\
   \end{array}\right)
   = U^T
    \left(\begin{array}{c}
     \tilde{A}_M^{\overline{6}} \\
     \tilde{A}_M^{\overline{9}} \\
     \tilde{A}_M^{\overline{10}} \\
     \tilde{A}_M^{\overline{16}} \\
     \tilde{A}_M^{\overline{19}} \\
     \tilde{A}_M^{\overline{20}} \\
     \tilde{A}_M^{18} \\
  \end{array}\right)  ,
\eea
and
$A_M^{q9}  =\tilde{A}_M^{9}$,
$A_M^{q10} =\tilde{A}_M^{10}$.
Here $U$ is given in Eq.~(\ref{defu}).

\subsection{$A_M^{qi}\leftrightarrow \tilde{A}_M^{%
\overline{i}}$}

\bea
 \left(\begin{array}{c}
   A_M^{q1} \\
   A_M^{q5} \\
   A_M^{q11} \\
   A_M^{q14} \\
   A_M^{q16} \\
   A_M^{q22} \\
   \end{array}\right)
   &\!\!\!=\!\!\!& {1\over \sqrt{2}}
   \left(\begin{array}{cccccc}
   {c +1\over 2} & {s\over \sqrt{2}} &
      -{c - 1\over 2}& {c + 1\over 2} 
    & {s\over \sqrt{2}}   &{c -1 \over 2} 
\\
 -{s\over \sqrt{2}} & c
   & {s\over \sqrt{2}} &  -{s\over \sqrt{2}}
  & c & -{s \over \sqrt{2}} 
\\
   -{c - 1\over 2} & {s\over \sqrt{2}} &
   {c + 1\over 2} &  {c-1 \over 2} 
  & -{s\over \sqrt{2}} & {c + 1\over 2}
\\
 -{s\over \sqrt{2}} & -c & {s\over \sqrt{2}} &
  {s\over \sqrt{2}} & c & {s\over \sqrt{2}}
\\
  {c-1 \over 2} & {s\over \sqrt{2}} &
   - {c+1 \over 2}  &  {c-1 \over 2} &
   {s\over \sqrt{2}} & {c + 1\over 2}
\\
 -{c + 1\over 2} & {s\over \sqrt{2}} &
   {c-1 \over 2} & {c + 1\over 2} &
  -{s\over \sqrt{2}} &  {c-1 \over 2}
\\   \end{array}\right)
    \left(\begin{array}{c}
     \tilde{A}_M^{\overline{1}} \\
     \tilde{A}_M^{\overline{4}} \\
     \tilde{A}_M^{\overline{7}} \\
     \tilde{A}_M^{\overline{11}} \\
     \tilde{A}_M^{\overline{14}} \\
     \tilde{A}_M^{\overline{17}} \\
  \end{array}\right)   ,
\\
%\eea
%\bea
  \left(\begin{array}{c}
    A_M^{q2} \\
    A_M^{q4} \\
    A_M^{q12} \\
    A_M^{q13} \\
    A_M^{q17} \\
    A_M^{q23} \\
    \end{array}\right)
   &\!\!\! =\!\!\!& {1\over \sqrt{2}}
   \left(\begin{array}{cccccc}
   {c +1\over 2} & -{s\over \sqrt{2}} &
      -{c - 1\over 2}& {c + 1\over 2} 
    & -{s\over \sqrt{2}}   &{c -1 \over 2} 
\\
 {s\over \sqrt{2}} & c
   & -{s\over \sqrt{2}} &  {s\over \sqrt{2}}
  & c & {s \over \sqrt{2}} 
\\
   -{c - 1\over 2} & -{s\over \sqrt{2}} &
   {c + 1\over 2} &  {c-1 \over 2} 
  & {s\over \sqrt{2}} & {c + 1\over 2}
\\
 {s\over \sqrt{2}} & -c & -{s\over \sqrt{2}} &
  -{s\over \sqrt{2}} & c & -{s\over \sqrt{2}}
\\
  {c-1 \over 2} & -{s\over \sqrt{2}} &
   - {c+1 \over 2}  &  {c-1 \over 2} &
   -{s\over \sqrt{2}} & {c + 1\over 2}
\\
 {c + 1\over 2} & {s\over \sqrt{2}} &
   -{c-1 \over 2} & -{c + 1\over 2} &
  -{s\over \sqrt{2}} &  -{c-1 \over 2}
\\   \end{array}\right)
     \left(\begin{array}{c}
     \tilde{A}_M^{\overline{2}} \\
     \tilde{A}_M^{\overline{3}} \\
     \tilde{A}_M^{\overline{8}} \\
     \tilde{A}_M^{\overline{12}} \\
     \tilde{A}_M^{\overline{13}} \\
     \tilde{A}_M^{\overline{18}} \\
  \end{array}\right)   ,
\\
%\eea
%\bea
  \left(\begin{array}{c}
   A_M^{q6} \\
   A_M^{q19} \\
   A_M^{q20} \\
   \end{array}\right)
  &\!\!\! =\!\!\!& {1\over \sqrt{2}}
   \left(\begin{array}{ccc}
   c & s & 1 \\
   -\sqrt{2} s & \sqrt{2} c& 0 \\
   -c & -s & 1 \\
   \end{array}\right) 
   \left(\begin{array}{c}
    \tilde{A}_M^{\overline{5}} \\
    \tilde{A}_M^{19} \\
    \tilde{A}_M^{\overline{15}} \\
    \end{array}\right) ,
  \quad
   \left(\begin{array}{c}
    A_M^{q3} \\
    A_M^{q7} \\
    A_M^{q8} \\
    A_M^{q15} \\
    A_M^{q18} \\
    A_M^{q21} \\
    A_M^{q24} \\
    \end{array}\right)
  = T 
     \left(\begin{array}{c}
     \tilde{A}_M^{\overline{6}} \\
     \tilde{A}_M^{\overline{9}} \\
     \tilde{A}_M^{\overline{10}} \\
     \tilde{A}_M^{\overline{16}} \\
     \tilde{A}_M^{\overline{19}} \\
     \tilde{A}_M^{\overline{20}} \\
     \tilde{A}_M^{18} \\
  \end{array}\right)  ,
\eea
and 
$A_M^{q9}  =\tilde{A}_M^{9}$,
$A_M^{q10} =\tilde{A}_M^{10}$. Here
\bea
  && T
 = {1\over \sqrt{2}} \times 
\nonumber
\\ 
  &&
    \left(\begin{array}{ccccccc}
    -{c s\over \sqrt{2}} &
  {405 + 438 c -11 c^2
  + (58  + 28 c
  +  10 c^2) \sqrt{10} \over 
  32(26 + 3 \sqrt{10})}  &
{1-c \over 2} & {-s\over \sqrt{2}} & 
   {3 + c^2\over 4} & {\sqrt{5} (1-c^2)\over 4}
 & {s^2\over 2 \sqrt{2}} 
\\
  2 c^2 -1 &
 -{(-9 \sqrt{2} + \sqrt{5} 
   + (\sqrt{2} - \sqrt{5}) c) s \over 16} &
 {-s\over \sqrt{2}} & c &
   {c s\over \sqrt{2}} &
   {-5 c s \over \sqrt{10}} & -c s
\\
 {5 c s \over  \sqrt{6}} &
 -{(c-1) (383 + 78 \sqrt{10}
   + (-55 + 50 \sqrt{10}) c)\over 
   32 \sqrt{3} (26 + 3 \sqrt{10})}
 &  
 {c +1\over 2 \sqrt{3}}
 & {s\over \sqrt{6}} 
 & {5 (1-c^2) \over 4 \sqrt{3}}
 & {5 (5 c^2 -1)\over 4\sqrt{15}}
 & -{5 s^2 \over 2 \sqrt{6}}
\\
 -{s \over 2 \sqrt{3}} &
 {1169 - 438 c - 11 c^2
  + (642  - 28 c  + 10 c^2) \sqrt{10}
  \over 32\sqrt{6}(26 + 3 \sqrt{10})}
  &
 {c+ 4\over  2 \sqrt{6}}
 & {c s\over 2 \sqrt{3}} 
 & {c^2-7 \over 4 \sqrt{6}}
 & {- 5 (c^2 +1) \over 4\sqrt{30}}
 & {s^2\over 4 \sqrt{3}}
\\
 \sqrt{2} c s &
{(-2 + \sqrt{10}) s^2\over 32}
 &
  0 & 0 & {s^2\over 2}
   & {-\sqrt{5} s^2\over 2}
   & {c^2 +1 \over \sqrt{2}}
\\ 
 1- 2 c^2 &
  {(9\sqrt{2} - \sqrt{5} + (\sqrt{2} - \sqrt{5}) c) s
  \over 16} 
  &
 {-s\over \sqrt{2}} & c
 & {-c s\over \sqrt{2}}
 & {5 cs\over \sqrt{10}} & c s
\\
 -{\sqrt{5} c s\over 4} &
 {3 (-5 + \sqrt{10}) + c (10 - 18 \sqrt{10}
  - (-5 + \sqrt{10}) c)\over 64} 
  &
   {\sqrt{10} c\over 4}  & {\sqrt{5} s\over 2}
 & {5 (c^2 +1)\over 4\sqrt{10}}
 & {3 - 5 c^2 \over 4 \sqrt{2}}
 & {\sqrt{5} s^2\over 4}
\\  
    \end{array}\right) 
        .
\nonumber
\\ \label{tdef}
\eea

\subsection{$A_M^{q\overline{i}}\leftrightarrow 
\tilde{A}_M^i$}

\bea
  \left(\begin{array}{c}
   A_M^{q\overline{1}} \\
   A_M^{q\overline{4}} \\
   A_M^{q\overline{7}} \\
   A_M^{q\overline{11}}\\
   A_M^{q\overline{14}} \\
   A_M^{q\overline{17}} \\
  \end{array}\right)
  &\!\!\!= \!\!\!& {1\over \sqrt{2}}
  \left(\begin{array}{cccccc}
    {1 + c\over 2} & {s\over \sqrt{2}} &
    {1 - c\over 2} & {s\over \sqrt{2}} &
   {c-1 \over 2} & -{1 + c\over 2} \\
  -{s\over \sqrt{2}} &  c & 
  -{s \over \sqrt{2}} &  -c &
   -{s\over \sqrt{2}} & -{s\over \sqrt{2}} \\
 {1 - c \over 2} & -{s\over \sqrt{2}} &
  {1 + c \over 2} &  -{s\over \sqrt{2}} &
   -{1 + c\over 2} &  {c-1 \over 2} \\
   {1 + c\over 2} & {s\over \sqrt{2}} &
   {c-1 \over 2} & -{s\over \sqrt{2}} & 
   {c-1 \over 2} & {1 + c\over 2} \\
  -{s\over \sqrt{2}} & c & {s\over \sqrt{2}} & 
  c & -{s\over \sqrt{2}} &
   {s\over \sqrt{2}} \\
  {c-1\over 2} & {s\over \sqrt{2}} &
   {1 + c \over 2} & -{s\over \sqrt{2}} &
  {1 + c \over 2} & {c-1 \over 2}\\
  \end{array}\right) 
   \left(\begin{array}{c}
     \tilde{A}_M^1 \\
     \tilde{A}_M^5 \\
     \tilde{A}_M^{11} \\
     \tilde{A}_M^{14} \\
     \tilde{A}_M^{16} \\
     \tilde{A}_M^{22} \\
     \end{array}\right) ,
\\
%\eea
%\bea
  \left(\begin{array}{c}
    A_M^{q\overline{2}} \\
    A_M^{q\overline{3}} \\
    A_M^{q\overline{8}} \\
    A_M^{q\overline{12}} \\
    A_M^{q\overline{13}} \\
    A_M^{q\overline{18}} \\
    \end{array}\right)
 &\!\!\! =\!\!\!& {1\over \sqrt{2}}
  \left( \begin{array}{cccccc}
  {1 + c\over 2} & -{s\over \sqrt{2}} &
  {1 - c\over 2} & -{s\over \sqrt{2}} &
    {c-1 \over 2} & {1 + c\over 2} \\
 {s\over \sqrt{2}} & c & {s\over \sqrt{2}} &
  -c & {s\over \sqrt{2}} &
   -{s\over \sqrt{2}} \\
  {1 - c\over 2} &  {s\over \sqrt{2}} &
  {1 + c\over 2} & {s\over \sqrt{2}} & 
  -{1 + c\over 2} & {1 - c \over 2} \\
  {1 + c \over 2} & -{s\over \sqrt{2}} &
   {c-1 \over 2} & {s\over \sqrt{2}} &
    {c-1\over 2} & -{1 + c\over 2} \\
  {s\over \sqrt{2}} & c& -{s\over \sqrt{2}} & 
  c& {s\over \sqrt{2}} & {s\over \sqrt{2}} \\
 {c-1 \over 2} & -{s\over \sqrt{2}} &
  {1 + c\over 2} & {s \over \sqrt{2}} &
   {1 + c\over 2} & {1 - c\over 2} \\
  \end{array}\right)
  \left(\begin{array}{c}
    \tilde{A}_M^2 \\
    \tilde{A}_M^4 \\
    \tilde{A}_M^{12} \\
    \tilde{A}_M^{13} \\
    \tilde{A}_M^{17} \\
    \tilde{A}_M^{23} \\
    \end{array}\right) ,
\\
%\eea
%\bea
   \left(\begin{array}{c}
     A_M^{q\overline{5}} \\
     A_M^{q 19} \\
     A_M^{q\overline{15}} \\
     \end{array}\right)
    &\!\!\! =\!\!\!& {1\over \sqrt{2}}
     \left(\begin{array}{ccc}
    c & \sqrt{2} s & -c \\
   -s & \sqrt{2} c & s \\
    1 & 0 & 1 \\
      \end{array}\right)
 \left(\begin{array}{c}
    \tilde{A}_M^6 \\
    \tilde{A}_M^{19} \\
    \tilde{A}_M^{20} \\
    \end{array}\right) ,
\\
%\eea
%\bea
  \left(\begin{array}{c}
     A_M^{q\overline{6}} \\
     A_M^{q\overline{9}} \\
     A_M^{q\overline{10}} \\
     A_M^{q\overline{16}} \\
     A_M^{q\overline{19}} \\
     A_M^{q\overline{20}} \\
     A_M^{q 18} \\
     \end{array}\right)
   &\!\!\!=\!\!\!& {1\over \sqrt{2}}
    J
   \left(\begin{array}{c}
    \tilde{A}_M^3 \\
    \tilde{A}_M^7 \\
    \tilde{A}_M^8 \\
    \tilde{A}_M^{15} \\
    \tilde{A}_M^{18} \\
    \tilde{A}_M^{21} \\
    \tilde{A}_M^{24} \\
    \end{array}\right) ,
    \qquad
    J=(J_1, J_2, J_3) ,
      \label{jdef}
\eea
and 
$A_M^{q9}  =\tilde{A}_M^{9}$,
$A_M^{q10} =\tilde{A}_M^{10}$. Here
\bea
 J_1 &\!\!\!=\!\!\!& \left(\begin{array}{cc}
  {c s\over \sqrt{2}} &  2 c^2 -1
  \\
  {16 (1 + c)\over 
  26 + 3 \sqrt{10}} & 
   -{16 \sqrt{2} s \over 26 + 3 \sqrt{10}} 
\\ 
-{-18 + \sqrt{10} 
  + 4 (2 + \sqrt{10}) c\over 26 + 3 \sqrt{10}} &
  {4 (11 \sqrt{2} + 20 \sqrt{5}) s\over 293} 
\\
 {s\over \sqrt{2}} &  c
\\
{633 + 96 \sqrt{10} + c (-246 + 96 \sqrt{10} + 293 c)
 \over 1172 } &
  -{(-6 + 3 \sqrt{10} + (26 + 3 \sqrt{10}) c) s \over 
   \sqrt{2}(26 + 3 \sqrt{10})}
\\
-{(1 + c) (-25  - 11 \sqrt{10} + 15 c + 
      13 \sqrt{10} c) \over 2\sqrt{2} 
  (26 + 3 \sqrt{10})} &
  {(-160 + 41 \sqrt{10} + 
     293 \sqrt{10} c) s \over 586} 
\\
   {s^2 \over 2 \sqrt{2}} &  c s 
\\ 
  \end{array}\right) ,
\\
%\eea 
%\bea
 J_2  &\!\!\!=\!\!\!& \left(\begin{array}{cc}
-{5 c s\over \sqrt{6}} & {c s \over 2 \sqrt{3}} 
\\
-{16 (c-1)\over \sqrt{3}(26 + 3 \sqrt{10})} &
  -{8 \sqrt{6} (c-4)\over 3(26 + 3 \sqrt{10})} 
\\
 {18 - \sqrt{10} + 4 (2 + \sqrt{10}) c
  \over \sqrt{3}(26 + 3 \sqrt{10})} &
 {4 (18 - \sqrt{10} + (2 + \sqrt{10}) c)
  \over \sqrt{6}(26 + 3 \sqrt{10})} 
\\
 -{s\over \sqrt{6}} & -{s\over 2 \sqrt{3}} 
\\
 -{(c-1) (118 + 21 \sqrt{10} 
  + 5 (26 + 3 \sqrt{10}) c)\over 
   4\sqrt{3} (26+ 3 \sqrt{10})} &
 {-230 + 3 \sqrt{10} + 
   6 (2 - \sqrt{10}) c 
  + (26 + 3 \sqrt{10}) c^2 \over
  4\sqrt{6} (26+ 3 \sqrt{10})}
\\
 {-5 (1 + 3 \sqrt{10}) 
  + (-10 + 2 \sqrt{10}) c 
  + 5 (15 + 13 \sqrt{10}) c^2 \over 
  2\sqrt{6} (26+ 3 \sqrt{10})} &
 {25 - 21 \sqrt{10} + 
     2 (-5 + \sqrt{10}) c 
  - (15 + 13 \sqrt{10}) c^2) \over
  4\sqrt{3} (26+ 3 \sqrt{10})} 
\\
  -{5 s^2 \over 2 \sqrt{6}} &
   {s^2 \over 4 \sqrt{3}}  
\\
  \end{array}\right) ,
\\
%\eea 
%\bea
  J_3 &\!\!\!=\!\!\!& \left(\begin{array}{ccc}
  0 & 1 - 2 c^2 & -{\sqrt{5} s\over 2}
\\
 0 & {4 \sqrt{2} (1 + 5 c) s \over 
  26 + 3 \sqrt{10}} &
  -{8 (-15 + 13 \sqrt{10}) c \over 293}
\\
  0 & -{(11 \sqrt{2} + 20 \sqrt{5}) (1 + 5 c) s
   \over 293} &
   {2 (100 + 11 \sqrt{10}) c\over 293}
\\
  -\sqrt{2} c s & c & 
   {\sqrt{5} c s \over 2}
\\
  {s^2\over 2} &
  -{(17 \sqrt{2} + 3 \sqrt{5}+ 7\sqrt{2} c - 3\sqrt{5} 
c) s \over 
   26 + 3 \sqrt{10}} &
 {293 \sqrt{10} + c (-960 + 246 \sqrt{10} 
  + 293 \sqrt{10} c) \over
  2344}
\\
 -{\sqrt{5} s^2 \over 2} &
  {(5 + 3 \sqrt{10} - (5 + 11 \sqrt{10}) c) s\over
  26 + 3 \sqrt{10}} &
  {3(26 + 3 \sqrt{10}) - 
   5 c (-4 + 2 \sqrt{10} 
 + (26 + 3 \sqrt{10}) c)\over 
  4\sqrt{2} (26 + 3 \sqrt{10})}
\\
  {1 + c^2 \over \sqrt{2}} & 
  cs & {\sqrt{5} s^2 \over 4}
\\
  \end{array}\right) .
\eea

\section{Mode expansion of gauge bosons \label{ap:gm}}

In the case where the rank-reducing boundary is located at
the Planck brane,
gauge bosons are expanded in terms of fundamental functions as
\bea
  \left(\begin{array}{c}
   A_\mu^{q\overline{1}} \\
   A_\mu^{q\overline{4}} \\
   A_\mu^{q\overline{7}} \\
   A_\mu^{q\overline{11}}\\
   A_\mu^{q\overline{14}} \\
   A_\mu^{q\overline{17}} \\
  \end{array}\right)
  &\!\!\!=\!\!\!&
  \left(\begin{array}{cccccc}
    {1 + c\over 2} & {s\over \sqrt{2}} &
    {1 - c\over 2} & {s\over \sqrt{2}} &
   {c-1 \over 2} & -{1 + c\over 2} \\
  -{s\over \sqrt{2}} &  c & 
  -{s \over \sqrt{2}} &  -c &
   -{s\over \sqrt{2}} & -{s\over \sqrt{2}} \\
 {1 - c \over 2} & -{s\over \sqrt{2}} &
  {1 + c \over 2} &  -{s\over \sqrt{2}} &
   -{1 + c\over 2} &  {c-1 \over 2} \\
   {1 + c\over 2} & {s\over \sqrt{2}} &
   {c-1 \over 2} & -{s\over \sqrt{2}} & 
   {c-1 \over 2} & {1 + c\over 2} \\
  -{s\over \sqrt{2}} & c & {s\over \sqrt{2}} & 
  c & -{s\over \sqrt{2}} &
   {s\over \sqrt{2}} \\
  {c-1\over 2} & {s\over \sqrt{2}} &
   {1 + c \over 2} & -{s\over \sqrt{2}} &
  {1 + c \over 2} & {c-1 \over 2}\\
  \end{array}\right) 
   \left(\begin{array}{c}
     C_{A,n}^1 C(z;\lambda_n) \\
     C_{A,n}^5 S(z;\lambda_n)\\
     C_{A,n}^{11} S(z;\lambda_n) \\
     C_{A,n}^{14} S(z;\lambda_n)\\
     C_{A,n}^{16} S(z;\lambda_n)\\
     C_{A,n}^{22} C(z;\lambda_n)\\
     \end{array}\right)
    {A_\mu^{(n)}(x) \over \sqrt{2}} ,
\nonumber
\\    
%\eea
%\bea
  \left(\begin{array}{c}
    A_\mu^{q\overline{2}} \\
    A_\mu^{q\overline{3}} \\
    A_\mu^{q\overline{8}} \\
    A_\mu^{q\overline{12}} \\
    A_\mu^{q\overline{13}} \\
    A_\mu^{q\overline{18}} \\
    \end{array}\right)
 &\!\!\! =\!\!\!&
  \left( \begin{array}{cccccc}
  {1 + c\over 2} & -{s\over \sqrt{2}} &
  {1 - c\over 2} & -{s\over \sqrt{2}} &
    {c-1 \over 2} & {1 + c\over 2} \\
 {s\over \sqrt{2}} & c & {s\over \sqrt{2}} &
  -c & {s\over \sqrt{2}} &
   -{s\over \sqrt{2}} \\
  {1 - c\over 2} &  {s\over \sqrt{2}} &
  {1 + c\over 2} & {s\over \sqrt{2}} & 
  -{1 + c\over 2} & {1 - c \over 2} \\
  {1 + c \over 2} & -{s\over \sqrt{2}} &
   {c-1 \over 2} & {s\over \sqrt{2}} &
    {c-1\over 2} & -{1 + c\over 2} \\
  {s\over \sqrt{2}} & c& -{s\over \sqrt{2}} & 
  c& {s\over \sqrt{2}} & {s\over \sqrt{2}} \\
 {c-1 \over 2} & -{s\over \sqrt{2}} &
  {1 + c\over 2} & {s \over \sqrt{2}} &
   {1 + c\over 2} & {1 - c\over 2} \\
  \end{array}\right)
  \left(\begin{array}{c}
    C_{A,n}^2 C(z;\lambda_n) \\
    C_{A,n}^4 S(z;\lambda_n)\\
    C_{A,n}^{12} S(z;\lambda_n)\\
    C_{A,n}^{13} S(z;\lambda_n)\\
    C_{A,n}^{17} S(z;\lambda_n)\\
    C_{A,n}^{23} C(z;\lambda_n)\\
    \end{array}\right) {A_\mu^{(n)}(x) \over \sqrt{2}} ,
\nonumber
\\
%\eea
%\bea
   \left(\begin{array}{c}
     A_\mu^{q\overline{5}} \\
     A_\mu^{q 19} \\
     A_\mu^{q\overline{15}} \\
     \end{array}\right)
   &\!\!\! =\!\!\!&   
     \left(\begin{array}{ccc}
    c & \sqrt{2} s & -c \\
   -s & \sqrt{2} c & s \\
    1 & 0 & 1 \\
      \end{array}\right)
 \left(\begin{array}{c}
    C_{A,n}^6 S(z;\lambda_n)\\
    C_{A,n}^{19} S(z;\lambda_n)\\
    C_{A,n}^{20} S(z;\lambda_n)\\
    \end{array}\right) {A_\mu^{(n)}(x)\over \sqrt{2}} , 
\nonumber
\\
%\eea
%\bea
  \left(\begin{array}{c}
     A_\mu^{q\overline{6}} \\
     A_\mu^{q\overline{9}} \\
     A_\mu^{q\overline{10}} \\
     A_\mu^{q\overline{16}} \\
     A_\mu^{q\overline{19}} \\
     A_\mu^{q\overline{20}} \\
     A_\mu^{q 18} \\
     \end{array}\right)
   &\!\!\!=\!\!\!& 
    J
   \left(\begin{array}{c}
    C_{A,n}^3 C(z;\lambda_n)\\
    C_{A,n}^7 S(z;\lambda_n)\\
    C_{A,n}^8 C(z;\lambda_n)\\
    C_{A,n}^{15} C(z;\lambda_n)\\
    C_{A,n}^{18} S(z;\lambda_n)\\
    C_{A,n}^{21} S(z;\lambda_n)\\
    C_{A,n}^{24} C(z;\lambda_n)\\
    \end{array}\right) {A_\mu^{(n)}(x) \over \sqrt{2}},
  \qquad    
  \left\{ \begin{array}{c}
   A_\mu^{q9}  = C_{A,n}^{9} S(z;\lambda_n) 
        A_\mu^{(n)}(x) ,
  \\
   A_\mu^{q10} = C_{A,n}^{10} S(z;\lambda_n)
       A_\mu^{(n)}(x) .
  \\     
   \end{array}\right.
\eea
where $J$ is given in Eq.~(\ref{jdef}).

In the case where the rank-reducing boundary is located at
the TeV brane,
gauge bosons are 
expanded in terms of fundamental functions as
\bea
 \left(\begin{array}{c}
   A_\mu^{q1} \\
   A_\mu^{q5} \\
   A_\mu^{q11} \\
   A_\mu^{q14} \\
   A_\mu^{q16} \\
   A_\mu^{q22} \\
   \end{array}\right)
  &\!\!\! = \!\!\!&
   \left(\begin{array}{cccccc}
   {c +1\over 2} & {s\over \sqrt{2}} &
      -{c - 1\over 2}& {c + 1\over 2} 
    & {s\over \sqrt{2}}   &{c -1 \over 2} 
\\
 -{s\over \sqrt{2}} & c
   & {s\over \sqrt{2}} &  -{s\over \sqrt{2}}
  & c & -{s \over \sqrt{2}} 
\\
   -{c - 1\over 2} & {s\over \sqrt{2}} &
   {c + 1\over 2} &  {c-1 \over 2} 
  & -{s\over \sqrt{2}} & {c + 1\over 2}
\\
 -{s\over \sqrt{2}} & -c & {s\over \sqrt{2}} &
  {s\over \sqrt{2}} & c & {s\over \sqrt{2}}
\\
  {c-1 \over 2} & {s\over \sqrt{2}} &
   - {c+1 \over 2}  &  {c-1 \over 2} &
   {s\over \sqrt{2}} & {c + 1\over 2}
\\
 -{c + 1\over 2} & {s\over \sqrt{2}} &
   {c-1 \over 2} & {c + 1\over 2} &
  -{s\over \sqrt{2}} &  {c-1 \over 2}
\\   \end{array}\right)
    \left(\begin{array}{c}
     C_{A,n}^{\overline{1}} C(z;\lambda_n) \\
     C_{A,n}^{\overline{4}} C(z;\lambda_n)\\
     C_{A,n}^{\overline{7}} C(z;\lambda_n)\\
     C_{A,n}^{\overline{11}} S(z;\lambda_n)\\
     C_{A,n}^{\overline{14}} S(z;\lambda_n)\\
     C_{A,n}^{\overline{17}} S(z;\lambda_n)\\
  \end{array}\right)  {A_\mu^{(n)}(x) \over \sqrt{2}} ,
\nonumber
\\
%\eea
%\bea
  \left(\begin{array}{c}
    A_\mu^{q2} \\
    A_\mu^{q4} \\
    A_\mu^{q12} \\
    A_\mu^{q13} \\
    A_\mu^{q17} \\
    A_\mu^{q23} \\
    \end{array}\right)
    &\!\!\! =\!\!\!&
   \left(\begin{array}{cccccc}
   {c +1\over 2} & -{s\over \sqrt{2}} &
      -{c - 1\over 2}& {c + 1\over 2} 
    & -{s\over \sqrt{2}}   &{c -1 \over 2} 
\\
 {s\over \sqrt{2}} & c
   & -{s\over \sqrt{2}} &  {s\over \sqrt{2}}
  & c & {s \over \sqrt{2}} 
\\
   -{c - 1\over 2} & -{s\over \sqrt{2}} &
   {c + 1\over 2} &  {c-1 \over 2} 
  & {s\over \sqrt{2}} & {c + 1\over 2}
\\
 {s\over \sqrt{2}} & -c & -{s\over \sqrt{2}} &
  -{s\over \sqrt{2}} & c & -{s\over \sqrt{2}}
\\
  {c-1 \over 2} & -{s\over \sqrt{2}} &
   - {c+1 \over 2}  &  {c-1 \over 2} &
   -{s\over \sqrt{2}} & {c + 1\over 2}
\\
 {c + 1\over 2} & {s\over \sqrt{2}} &
   -{c-1 \over 2} & -{c + 1\over 2} &
  -{s\over \sqrt{2}} &  -{c-1 \over 2}
\\   \end{array}\right)
     \left(\begin{array}{c}
     C_{A,n}^{\overline{2}}C(z;\lambda_n) \\
     C_{A,n}^{\overline{3}}C(z;\lambda_n) \\
     C_{A,n}^{\overline{8}}C(z;\lambda_n) \\
     C_{A,n}^{\overline{12}}S(z;\lambda_n) \\
     C_{A,n}^{\overline{13}}S(z;\lambda_n) \\
     C_{A,n}^{\overline{18}}S(z;\lambda_n) \\
  \end{array}\right) {A_\mu^{(n)}(x) \over \sqrt{2}} ,
\nonumber
\\
%\eea
%\bea
  \left(\begin{array}{c}
   A_\mu^{q6} \\
   A_\mu^{q19} \\
   A_\mu^{q20} \\
   \end{array}\right)
  &\!\!\! =\!\!\!&
   \left(\begin{array}{ccc}
   c & s & 1 \\
   -\sqrt{2} s & \sqrt{2} c& 0 \\
   -c & -s & 1 \\
   \end{array}\right) 
   \left(\begin{array}{c}
    C_{A,n}^{\overline{5}} C(z;\lambda_n) \\
    C_{A,n}^{19} S(z;\lambda_n) \\
    C_{A,n}^{\overline{15}} S(z;\lambda_n) \\
    \end{array}\right) {A_\mu^{(n)}(x)\over \sqrt{2}} ,
\nonumber
\\
%\eea
%\bea
  \left(\begin{array}{c}
    A_\mu^{q3} \\
    A_\mu^{q7} \\
    A_\mu^{q8} \\
    A_\mu^{q15} \\
    A_\mu^{q18} \\
    A_\mu^{q21} \\
    A_\mu^{q24} \\
    \end{array}\right)
  &\!\!\! = \!\!\!& T 
     \left(\begin{array}{c}
     C_{A,n}^{\overline{6}} C(z;\lambda_n)\\
     C_{A,n}^{\overline{9}} C(z;\lambda_n)\\
     C_{A,n}^{\overline{10}} C(z;\lambda_n)\\
     C_{A,n}^{\overline{16}} S(z;\lambda_n)\\
     C_{A,n}^{\overline{19}} S(z;\lambda_n)\\
     C_{A,n}^{\overline{20}} S(z;\lambda_n)\\
     C_{A,n}^{18} S(z;\lambda_n)\\
  \end{array}\right) A_\mu^{(n)}(x)  ,
 \qquad
   \left\{ \begin{array}{c}
   A_\mu^{q9}  = C_{A,n}^{9} S(z;\lambda_n) 
        A_\mu^{(n)}(x) ,
  \\
   A_\mu^{q10} = C_{A,n}^{10} S(z;\lambda_n)
       A_\mu^{(n)}(x) ,
  \\
    \end{array}\right.
\eea
where $T$ is given in Eq.~(\ref{tdef}).

\section{Matrices for 
Planck-brane boundary conditions \label{ap:bcpl}}

The matrix $E$ employed in Eq.~(\ref{plpl}) is defined as
$E=(E_1,E_2,E_3)$ with
\bea
 E_1 &\!\!\!=\!\!\!& \left(\begin{array}{cc}
  {c s\over \sqrt{2}} C' &  (2 c^2 -1) S'
  \\
  {16 (1 + c)\over 
  26 + 3 \sqrt{10}} C' & 
   -{16 \sqrt{2} s \over 26 + 3 \sqrt{10}} S' 
\\ 
-{-18 + \sqrt{10} 
  + 4 (2 + \sqrt{10}) c\over 26 + 3 \sqrt{10}} C'&
  {4 (11 \sqrt{2} + 20 \sqrt{5}) s\over 293} S'
\\
 {s\over \sqrt{2}} C&  c S
\\
{633 + 96 \sqrt{10} + c (-246 + 96 \sqrt{10} + 293 c)
 \over 1172 } C&
  -{(-6 + 3 \sqrt{10} + (26 + 3 \sqrt{10}) c) s \over 
   \sqrt{2}(26 + 3 \sqrt{10})} S
\\
-{(1 + c) (-25  - 11 \sqrt{10} + 15 c + 
      13 \sqrt{10} c) \over 2\sqrt{2} 
  (26 + 3 \sqrt{10})} C&
  {(-160 + 41 \sqrt{10} + 
     293 \sqrt{10} c) s \over 586} S 
\\
   {s^2 \over 2 \sqrt{2}} C &  c s S 
\\ 
  \end{array}\right) ,
\nonumber
\\
%\eea 
%\bea
 E_2  &\!\!\!=\!\!\!& \left(\begin{array}{cc}
-{5 c s\over \sqrt{6}} C'& {c s \over 2 \sqrt{3}} C'
\\
-{16 (c-1)\over \sqrt{3}(26 + 3 \sqrt{10})} C'&
  -{8 \sqrt{6} (c-4)\over 3(26 + 3 \sqrt{10})} C'
\\
 {18 - \sqrt{10} + 4 (2 + \sqrt{10}) c
  \over \sqrt{3}(26 + 3 \sqrt{10})} C'&
 {4 (18 - \sqrt{10} + (2 + \sqrt{10}) c)
  \over \sqrt{6}(26 + 3 \sqrt{10})} C'
\\
 -{s\over \sqrt{6}} C& -{s\over 2 \sqrt{3}} C
\\
 -{(c-1) (118 + 21 \sqrt{10} 
  + 5 (26 + 3 \sqrt{10}) c)\over 
   4\sqrt{3} (26+ 3 \sqrt{10})} C&
 {-230 + 3 \sqrt{10} + 
   6 (2 - \sqrt{10}) c 
  + (26 + 3 \sqrt{10}) c^2 \over
  4\sqrt{6} (26+ 3 \sqrt{10})} C
\\
 {-5 (1 + 3 \sqrt{10}) 
  + (-10 + 2 \sqrt{10}) c 
  + 5 (15 + 13 \sqrt{10}) c^2 \over 
  2\sqrt{6} (26+ 3 \sqrt{10})} C&
 {25 - 21 \sqrt{10} + 
     2 (-5 + \sqrt{10}) c 
  - (15 + 13 \sqrt{10}) c^2) \over
  4\sqrt{3} (26+ 3 \sqrt{10})} C
\\
  -{5 s^2 \over 2 \sqrt{6}} C&
   {s^2 \over 4 \sqrt{3}}  C
\\
  \end{array}\right) ,
\nonumber
\\
%\eea 
%\bea
  E_3 &\!\!\!=\!\!\!& \left(\begin{array}{ccc}
  0 & (1 - 2 c^2)S' & -{\sqrt{5} s\over 2} C'
\\
 0 & {4 \sqrt{2} (1 + 5 c) s \over 
  26 + 3 \sqrt{10}} S'&
  -{8 (-15 + 13 \sqrt{10}) c \over 293} C'
\\
  0 & -{(11 \sqrt{2} + 20 \sqrt{5}) (1 + 5 c) s
   \over 293} S'&
   {2 (100 + 11 \sqrt{10}) c\over 293} C'
\\
  -\sqrt{2} c s S& c S& 
   {\sqrt{5} c s \over 2} C
\\
  {s^2\over 2} S&
  -{(17 \sqrt{2} + 3 \sqrt{5}+ 7\sqrt{2} c - 3\sqrt{5} 
c) s \over 
   26 + 3 \sqrt{10}} S&
 {293 \sqrt{10} + c (-960 + 246 \sqrt{10} 
  + 293 \sqrt{10} c) \over
  2344} C
\\
 -{\sqrt{5} s^2 \over 2} S&
  {(5 + 3 \sqrt{10} - (5 + 11 \sqrt{10}) c) s\over
  26 + 3 \sqrt{10}} S&
  {3(26 + 3 \sqrt{10}) - 
   5 c (-4 + 2 \sqrt{10} 
 + (26 + 3 \sqrt{10}) c)\over 
  4\sqrt{2} (26 + 3 \sqrt{10})} C
\\
  {1 + c^2 \over \sqrt{2}} S& 
  cs S& {\sqrt{5} s^2 \over 4} C
\\
  \end{array}\right) .
\nonumber
\\
\eea

The matrix $L$ employed in Eq.~(\ref{pltev}) is defined as
$L=(L_1 ~ L_2)$ with
\bea
  L_1 &\!\!\!=\!\!\!&  
    \left(\begin{array}{cc}
    -{c s\over \sqrt{2}} C' &
  {405 + 438 c -11 c^2
  + (58  + 28 c
  +  10 c^2) \sqrt{10} \over 
  32(26 + 3 \sqrt{10})}  C'
\\
   (2 c^2 -1) C & 
  -{(-9 \sqrt{2} + \sqrt{5} 
   + (\sqrt{2} - \sqrt{5}) c) s \over 16} C
\\
   {5 c s \over  \sqrt{6}} C' 
  & -{(c-1) (383 + 78 \sqrt{10}
   + (-55 + 50 \sqrt{10}) c)\over 
   32 \sqrt{3} (26 + 3 \sqrt{10})}
 C'
\\
   -{s \over 2 \sqrt{3}} C' 
  & {1169 - 438 c - 11 c^2
  + (642  - 28 c  + 10 c^2) \sqrt{10}
  \over 32\sqrt{6}(26 + 3 \sqrt{10})}
  C'
\\
  \sqrt{2} c s C & {(-2 + \sqrt{10}) s^2\over 32}
 C
\\ 
  (1- 2 c^2) C
  & {(9\sqrt{2} - \sqrt{5} + (\sqrt{2} - \sqrt{5}) c) s
  \over 16} 
  C
\\
  -{\sqrt{5} c s\over 4} C'
  & {3 (-5 + \sqrt{10}) + c (10 - 18 \sqrt{10}
  - (-5 + \sqrt{10}) c)\over 64} 
  C'
\\  
    \end{array}\right) ,
\nonumber
\\        
%\eea
%\bea
  L_2 &\!\!\!=\!\!\!&  
    \left(\begin{array}{ccccc}
{1-c \over 2} C'& {-s\over \sqrt{2}} S'& 
   {3 + c^2\over 4} S'& {\sqrt{5} (1-c^2)\over 4}
 S'& {s^2\over 2 \sqrt{2}} S'
\\
 {-s\over \sqrt{2}} C& c S&
   {c s\over \sqrt{2}} S&
   {-5 c s \over \sqrt{10}} S& -c s S
\\
 {c +1\over 2 \sqrt{3}}
 C'& {s\over \sqrt{6}} 
 S'& {5 (1-c^2) \over 4 \sqrt{3}}
 S'& {5 (5 c^2 -1)\over 4\sqrt{15}}
 S'& -{5 s^2 \over 2 \sqrt{6}} S'
\\
 {c+ 4\over  2 \sqrt{6}}
 C'& {c s\over 2 \sqrt{3}} 
 S'& {c^2-7 \over 4 \sqrt{6}}
 S'& {- 5 (c^2 +1) \over 4\sqrt{30}}
 S'& {s^2\over 4 \sqrt{3}} S'
\\
  0 & 0 & {s^2\over 2}
   S& {-\sqrt{5} s^2\over 2}
   S& {c^2 +1 \over \sqrt{2}} S
\\ 
 {-s\over \sqrt{2}} C& c
 S& {-c s\over \sqrt{2}}
 S& {5 cs\over \sqrt{10}} S& c s S
\\
   {\sqrt{10} c\over 4}  C'& {\sqrt{5} s\over 2}
 S'& {5 (c^2 +1)\over 4\sqrt{10}}
 S'& {3 - 5 c^2 \over 4 \sqrt{2}}
 S'& {\sqrt{5} s^2\over 4} S'
\\  
    \end{array}\right) . 
\nonumber
\\     
\eea

\section{A determination of mode functions \label{ap:wmode}}

In this appendix, we give a derivation for
determining the coefficients 
for the mode functions for a real part of $W$ boson
by solving Eq.~(\ref{wbc}).
Eq.~(\ref{wbc}) is simplified as
\bea
   0=
   \left(\begin{array}{cccccc}
   (c +1) C' & 0 &
      (1 -c) C' & 0 
    & \sqrt{2} s S'   &0
\\
  0 & \sqrt{2} c C
   & 0 &  -s S
  & 0& -s S
\\
   (1-c) C & 0 &
   (c + 1) C&  0 
  & -\sqrt{2} s S& 0
\\
 -s C& 0 & s C &
  0& \sqrt{2} c S& 0
\\
  0& \sqrt{2} s C&
   0 &  (c-1) S&
   0& (c + 1) S
\\
  0& \sqrt{2} s C'&
   0& (c + 1) S'&
   0&  (c-1) S'
\\   \end{array}\right)
    \left(\begin{array}{c}
     C_{A,n}^{\overline{1}} \\
     C_{A,n}^{\overline{4}} \\
     C_{A,n}^{\overline{7}} \\
     C_{A,n}^{\overline{11}} \\
     C_{A,n}^{\overline{14}} \\
     C_{A,n}^{\overline{17}} \\
  \end{array}\right) .
  \nonumber
 \\
   \label{solhw}
\eea
In this expression, it is seen that 
the components ($C_{A,n}^{\overline{1}}$, $C_{A,n}^{\overline{7}}$,
$C_{A,n}^{\overline{14}}$)
and the components ($C_{A,n}^{\overline{4}}$,
$C_{A,n}^{\overline{11}}$,
$C_{A,n}^{\overline{17}}$)
are decoupled.
Therefore we can calculate each set of the components 
independently.
A part of the components ($C_{A,n}^{\overline{1}}$, 
$C_{A,n}^{\overline{7}}$,
$C_{A,n}^{\overline{14}}$) in Eq.~(\ref{solhw}) 
are given by
\bea
 &&   (c+1) C' C_{A,n}^{\overline{1}}
    +(1-c) C' C_{A,n}^{\overline{7}}
     +\sqrt{2}s S' C_{A,n}^{\overline{14}} 
     =0 ,
     \label{weq1}
\\
  &&
   (1-c) C C_{A,n}^{\overline{1}}
     +(c+1) C C_{A,n}^{\overline{7}}
      -\sqrt{2} s S C_{A,n}^{\overline{14}} =0 ,
  \label{weq2}
\\
  && -s C C_{A,n}^{\overline{1}}
    +s C C_{A,n}^{\overline{7}}
     +\sqrt{2} c S C_{A,n}^{\overline{14}} =0 .
  \label{weq3}
\eea
From Eqs.~(\ref{weq1}) and (\ref{weq2}),
$C_{A,n}^{\overline{7}}$ and $C_{A,n}^{\overline{14}}$
are obtained as
\bea
  C_{A,n}^{\overline{7}}
    =-{2C'S +\lambda (1-c)\over
    2C'S +\lambda (1+c)} C_{A,n}^{\overline{1}} ,
\qquad
  C_{A,n}^{\overline{14}}
   = -{\sqrt{2} c\over s}
     {2CC'\over
     2C'S +\lambda(1+c)}
      C_{A,n}^{\overline{1}} .
\eea
The left-hand side in Eq.~(\ref{weq3}) is
\bea
  -{2C\over s} {2C'S +\lambda s^2\over
    2C'S +\lambda (1+c)} C_{A,n}^{\overline{1}} .
\eea
This equation vanishes for $W$ boson whose mass 
eigenvalue equation is 
$2C'S =-\lambda_W s^2$.
For $\lambda=\lambda_W$, the coefficients
$C_{A,n}^{\overline{7}}$ and $C_{A,n}^{\overline{14}}$ are
\bea
   C_{A,n}^{\overline{7}} 
    ={1-c\over 1+c}C_{A,n}^{\overline{1}} ,
    \qquad
  C_{A,n}^{\overline{14}}
   ={\sqrt{2} s\over 1+c}
    {C\over S} C_{A,n}^{\overline{1}} .
     \label{wcoef}
\eea
A part of the components
($C_{A,n}^{\overline{4}}$,
$C_{A,n}^{\overline{11}}$,
$C_{A,n}^{\overline{17}}$) in Eq.~(\ref{solhw}) are given by
\bea
  && \sqrt{2} cC C_{A,n}^{\overline{4}}
   -sS C_{A,n}^{\overline{11}}
   - sS C_{A,n}^{\overline{17}} =0 ,
   \label{weq2-1}
\\
 && \sqrt{2} s C C_{A,n}^{\overline{4}}
  +(c-1)S  C_{A,n}^{\overline{11}}
  +(c+1) S C_{A,n}^{\overline{17}} =0 ,
  \label{weq2-2}
\\
 &&  \sqrt{2} sC' C_{A,n}^{\overline{4}}
  +(c+1) S' C_{A,n}^{\overline{11}}
  +(c-1) S' C_{A,n}^{\overline{17}} =0 .
   \label{weq2-3}
\eea
From Eqs.~(\ref{weq2-2}) and (\ref{weq2-3}),
$C_{A,n}^{\overline{4}}$ and $C_{A,n}^{\overline{17}}$ are
obtained as
\bea
  C_{A,n}^{\overline{4}}
   = -{\sqrt{2} c\over s}
     {2SS'\over 2C'S +(1-c)\lambda}
       C_{A,n}^{\overline{11}}
 \qquad
 C_{A,n}^{\overline{17}}
  ={2C'S + \lambda (1+c)\over
   2C'S +\lambda (1-c)} C_{A,n}^{\overline{11}}  .
\eea   
The left-hand side in Eq.~(\ref{weq2-1}) is
\bea
   -{2S\over s} {2C'S + \lambda(1+c^2)
 \over 2C'S +\lambda(1-c)}  C_{A,n}^{\overline{11}} .
\eea
The condition that this equation vanishes 
means $C_{A,n}^{\overline{11}}=0$.
Combining this with Eq.~(\ref{wcoef}), 
we obtain a real component of
$W$ boson as Eq.~(\ref{wmodeh}).

\end{appendix}

\newpage

%\vspace*{10mm}
%%%%%%%%%%%%% BIBLIOGRAPHY (US) %%%%%%%%%%%%%%%%%%%%

%%%%%%%%%%%%%%%%%%%%%

\end{document}